\documentclass[12pt,british,12pt, a4paper,twoside,openright]{article}

\usepackage[OT1,T1]{fontenc}
\usepackage[utf8]{inputenc}
\usepackage{babel}
\usepackage{textcomp}
\usepackage{amsmath}
\usepackage{amsthm}
\usepackage{amssymb}
\usepackage[authoryear]{natbib}
\usepackage[unicode=true]{hyperref}

\makeatletter

\providecommand{\tabularnewline}{\\}

\usepackage[LGR,T1,OT1]{fontenc}
\newcommand{\textgreek}[1]{\begingroup\fontencoding{LGR}\selectfont#1\endgroup}

\usepackage{yfonts}
\newcommand{\ts}[1]{\textswab{#1}}
\usepackage[unicode=true]{hyperref}

\usepackage{latexsym}
\usepackage{indentfirst}
\usepackage{graphicx}
\usepackage{amsfonts,amstext,amscd}
\usepackage{mathrsfs}
\usepackage{fancyhdr}
\usepackage{times}
\usepackage{enumerate}
\usepackage{natbib}
\usepackage[hang,flushmargin]{footmisc}

\newtheorem{lem}{Lemma}[section]

\newtheorem{thm}{Theorem}[section]

\newtheorem{coro}{Corollary}[section]

\fancypagestyle{plain}{%
\fancyhf{}%
\fancyhead[LO]{\textbf{Science \& Philosophy}
}
\fancyhead[RO]{Volume 12(2), 2024 }

}

\makeatother

\begin{document}
\title{The construction of quantum mechanics from electromagnetism. Theory
and hydrogen atom }
\author{Hernán Gustavo Solari\thanks{Departamento de Física, FCEN-UBA and IFIBA-CONICET; Pabellón I, Ciudad
Universitaria (1428) - C.A.B.A - Argentina. email: solari@df.uba.ar
Orcid: 0000-0003-4287-1878} {\ } and Mario Alberto Natiello\thanks{Centre for Mathematical Sciences, Lund University. Box 118, S 221
00 LUND, Sweden. email: mario.natiello@math.lth.se (corresponding
author) Orcid: 0000-0002-9481-7454}}
\date{}
\maketitle
\begin{abstract}
We reconstruct Quantum Mechanics in a way that harmonises with classical
mechanics and electromagnetism, free from mysteries or paradoxes as
the \emph{collapse of the wave-function} or \emph{Schrödinger's cat.}
The construction is inspired by de Broglie's and Schrödinger's wave
mechanics while the unifying principle is Hamilton's principle of
least action, which separates natural laws from particular circumstances
such as initial conditions and leads to the conservation of energy
for isolated systems. 

In Part I we construct the Quantum Mechanics of a charged unitary
entity and prescribe the form in which the entity interacts with other
charged entities and matter in general. In Part II we address the
quantum mechanics of the hydrogen atom, testing the correctness and
accuracy of the general description. The relation between electron
and proton in the atom is described systematically in a construction
that is free from analogies or ad-hoc derivations and it supersedes
conventional Quantum Mechanics (whose equations linked to measurements
can be recovered). We briefly discuss why the concept of isolation
built in Schrödinger's time evolution is not acceptable and how it
immediately results in the well known measurement paradoxes of quantum
mechanics. We also discuss the epistemic grounds of the development
and provide a criticism of instrumentalism, the leading philosophical
perspective behind conventional Quantum Mechanics. 
\end{abstract}
\textbf{Keywords:}\textsf{ }Wave mechanics; Schrödinger's equation;
Atomic energy levels; critical epistemology; instrumentalism; consilience;
unity of thought; epistemic praxis; \\
\textbf{AMS subject classification:} 81P05, 81Q05, 81V45\footnote{Received on November 14, 2024. Accepted on December 30, 2024. Published
on December 31, 2024. DOI: 10.23756/sp.v12i2.1659. ISSN 2282-7757;
eISSN 2282-7765. ©Natiello and Solari. This paper is published under
the CC-BY licence agreement.}

\newpage{}

\part*{Introduction }

Most physicists consider their science deals with the discovery of
the laws of nature. However, during the early years of the XX century
a most prominent group of physicists completely dropped this position
and sustained that physics was merely a collection of useful formulae
(see for example \citep{krag90}). Both groups of physicists would
however agree that the relevant product of physics was the collection
of formulae or plans which allowed them to make predictions. Actually,
successful predictions, and insights, boosted technological applications
justifying in front of the whole society the liberties granted to
scientists. In short, predictive success was the key of the game called
Science since late in the XIX century.

Under this philosophical perspective, called \emph{instrumentalism}
(a form of \emph{utilitarianism}), the process of construction of
theories --including their grounding-- becomes less relevant and
it is usually claimed to lie outside the science (for example in \citep{popp59})
or is substituted by a narrative that allows students to accept the
desired formulae with lesser intellectual effort (sometimes called
``didactic transposition''). Consider as an example the widespread
misconception about absolute space being fundamental for Newtonian
mechanics. However, the philosophy that developed science since (at
least) Galileo Galilei and until the second industrial revolution
(around 1870) was rationalist and considered the rational construction
of a theory a most relevant element in judging its quality. For the
very few in physics that recognise that knowledge represents the encounter
of the subject and the object and as such it carries elements of both
of them, the rational construction of theories remains important and
it is possible to suspect that, if we turn to the old philosophy of
Science, some of the utilitarian theories would show defects or imperfections
that demand attention.

A long-term project in this direction has been under development during
the last ten years reconstructing Mechanics \citep{sola18b} (and
particularly recovering the concept of ``inertial system'' much
distorted in textbooks \citep{sola22}), making explicit the dualist
epistemology and the relevance of phenomenology for science \citep{Solari22-phenomenologico}
and furthermore, rescuing a few rules of reason that pervade the old
construction of Science \citep{sola23-razo-retro}. A second constructive
step consisted in developing electromagnetism in terms of relational
space \citep{solari2022symmetries}, thus reclaiming Gauss' rational
project, which was left aside in favour of the idea of the ether and
subsequently forgotten under the utilitarian perspective. Having recovered
an Electromagnetism that is in harmony with Classical Mechanics, we
present now the construction of a Quantum Mechanics that is in harmony
with Classical Mechanics and Electromagnetism.

\subsection*{Background}

Under the instrumentalist guide, the period 1870-1930 witnessed the
consolidation of Electromagnetic theory (EM) and the birth of Quantum
Mechanics (QM). The success of both theories as predictive tools (and
thus as mediators of technological development) can hardly be challenged.
The interpretative part is often modelled resorting to mechanical
analogies \citep{mach12-popular,boltzmann1974philosphical}, although
a conflict arose with Electromagnetism since a mechanical (Galilean)
framework and the constancy of the velocity of light appeared to be
incompatible. The underlying model in Bohr's atomic theory is also
mechanical, with a planetary electron revolving around the atomic
nucleus. A conflict between theories becomes evident since an accelerated
charge is expected to radiate, thus losing energy and collapsing towards
the nucleus. The solution to the conflict was to negate it: In the
atomic realm radiation is not present and electronic orbits are stable.
In Bohr's atomic model, stable orbits have to satisfy a quantisation
condition (that the angular momentum is an integer multiple of Planck's
constant, $\hbar$) and electrons can only undergo transitions among
these stable orbits. Nothing is said in the model about the existence
of other states. Later, the Schrödinger model will assume that the
possible states of the atom are linear combinations of stationary
states and stable as well. In both models all possible atomic states
are stable and radiation is suppressed. However, it is not possible
to ascertain which is the state of the atom since measurements make
the state to ``collapse'' into a stationary (measurable) energy
state. This dogma is often indoctrinated by resource to Schrödinger's
\emph{cat paradox}. It would appear that the laws of EM are then different
for microscopic and macroscopic distances. The possibility of encompassing
EM and QM in one broader theory is hopeless in their currently accepted
shape since EM and QM are contradictory in their foundations.

The EM that is acceptable for atomic physics displays a number of
phenomena that are very difficult to explain. Consider e.g., spontaneous
emission (overwhelmingly verified in Laser physics \citep{carm99,chow94})
or the reduction postulate of QM transforming the state of the system
at the moment of measurement. A natural question arises: At what distance
is it mandatory to change theories? When and how EM ceases being applicable
and QM is a must? Are there other experimentally observable consequences?
The theories now rest on elements which are not amenable to experimental
examination and moreover, there are questions that cannot be asked. 

EM also suffers instrumentalist constraints on its own. The current
EM approach is rooted in the heritage of Hertz, being his preferred
interpretation that the EM fields are associated to space (the ether
for Hertz) and that forces surround matter \footnote{The expression is rather mysterious, but it can be found also in Faraday
(closer to reject it than to accept it) and in Maxwell (closer to
accept it than to reject it).}. Thus matter and EM fields have different ontogenesis. There exists
``empty space'', populated by fields (and subsequently by springs,
strings and branes), along with a space occupied by matter (originally
impenetrable). The realm of matter and with it atomic physics is divorced
from the space where fields dwell (despite the field conception being
rooted in the idea of force per unit charge). The link between both
realms is given mostly by the Lorentz force, which appears as an addendum
to Maxwell's equations displaying the Poincaré-Lorentz symmetry \citep{poin06b},
after repeated drift in the meaning of the symbols. The velocity,
that was relational in Maxwell and in experiments, became velocity
relative to the ether and later relative to an inertial frame \citep{assi94}.
The last two velocities have never been measured. For example, when
analysing experiments concerning Doppler effect, the standard theory
refers to a velocity with respect to an inertial system, but the measured
velocity is a relative one. We will not dwell in these issues, the
impossibility to physically contrast Special Relativity has been discussed
by \citep{essen1971special,essen1978relativity}, while its logical
inconsistency when regarded from an epistemology that encompasses
the phenomenological moment has been recently shown \citep{Solari22-phenomenologico}.
Strictly speaking, EM floats in a realm of ideas partially divorced
from observations.

The possibility of unifying EM and QM under a conception that cleaves
them in two separate entities became only possible by imposing the
quantification of imagined entities that populate ``empty space''.
This is the presently accepted approach, although their original proposers
finally rejected it and despite their internal inconsistencies that
would force the collapse of the underlying mathematical structure
\citep{nati15}. 

Unfortunately, alternative approaches have little or no room in the
scientific discussion. If the outcome in formulae of two theories
is practically the same; instrumentalism may be inclined to choose
the option with (apparently) less complications. Coming back to Newtonian
mechanics, if the outcome elaborated under the assumption of absolute
space to sustain the theory is the same as (a subset of) that developed
under Newton's concept of ``true motion'' \citep{newt87,sola22}
instrumentalists may find no benefit in presenting the second --more
abstract-- theory, to the price of accepting a conflicting concept
that cannot be demonstrated\footnote{Along the same lines, textbooks do not hesitate in presenting a simple
but incorrect derivation of the spin-orbit interaction along with
an ad-hoc correction called Thomas precession, instead of using the
approach that they consider ``correct'' \citep[p. 60]{kastberg2020structure}.}. We promptly recognise that in the academic formation of technical
professionals the difference is usually not very problematic, while
in the formation of experimental philosophers (or natural philosophers),
to decline the historical cognitive truth for instrumentalist reasons
appears as completely inappropriate and a safe source of future conflicts
\footnote{The distinction between the formation of professionals and philosophers
is in \citep{kant98}.}.

Instrumentalism is a weakened version of the phenomenological conception
presented in \citep{Solari22-phenomenologico}. This latter approach
is rooted in the work of C. Peirce \citep{peir94}, W. Whewell \citep{whewell1840philosophy}
and J. Goethe \citep{goethe09} developed under the influence of rational
humanism and nurtured by ideas of I. Kant \citep{kant87} and G. Hegel
\citep{hege01}. 

The ideas that allow us to traverse alternative paths were presented
a long time ago but their development was abandoned half way through.
Faraday observed that the more familiar ideas that privilege the concept
of matter before the concept of action could be confronted with the
approach that recognises that only actions can be detected while matter
is subsequently inferred, and as such exposed to the errors belonging
in conjectured entities. In this sense, Faraday recognises that matter
can only be conceived in relation to action and what we call matter
is the place where action is more intense \footnote{We read in \citep[p.447]{fara55}:
\begin{quote}
You are aware of the speculation which I some time since uttered respecting
that view of the nature of matter which considers its ultimate atoms
as centres of force, and not as so many little bodies surrounded by
forces, the bodies being considered in the abstract as independent
of the forces and capable of existing without them. In the latter
view, these little particles have a definite form and a certain limited
size; in the former view such is not the case, for that which represents
size may be considered as extending to any distance to which the lines
of force of the particle extend: the particle indeed is supposed to
exist only by these forces, and where they are it is.
\end{quote}
}. Such matter extends to infinity and is interpenetrable. Faraday's
vibrating rays theory eliminates the need to conceive an ether. The
relational electromagnetism initially developed by W. Weber \citep{assi94}
under Gauss influence, was later developed by the Göttingen school
and by L. Lorenz \citep{lore67} introducing Gauss' concept of retarded
action at distance. Maxwell expressed his surprise when a theory that
was based in principles very different from his could provide a body
of equations in such high accordance \citep{nive03}. From the Hertzian
point of view both theories are basically ``the same''. However,
when experiments later demonstrated the absence (nonexistence) of
a material ether \citep{michelson1887relative} the rational approach
of Faraday-Gauss-Lorenz was already forgotten, probably because the
epistemological basis of science had moved towards instrumentalism.

The development of Quantum Mechanics continued from Bohr's model into
Heisenberg's matrix computations, a completely instrumental approach
devoid of a priori meaning \citep{born55-nobel}. While the ideas
of de Broglie \citep{debroglie23,de1924recherches} and Schrödinger
\citep{schrodinger1926undulatory} revolutionised the field, at the
end of the Fifth Solvay conference (1927) the statistical interpretation
of quantum theory collected most of the support \citep{schr95}. It
is said that only Einstein and Schrödinger were not satisfied with
it. Einstein foresaw conflicts with Relativity theory, later made
explicit in the Einstein-Podolsky-Rosen paper \citep{eins35}. Schrödinger,
after several years of meditation and several changes of ideas, came
to the conclusion that the statistical interpretation was forced by
the, not properly justified, assumption of electrons and quantum particles
being represented by mathematical points \citep{schr95}. Certainly,
Schrödinger knew well that no statistical perspective was introduced
in his construction of ondulatory mechanics. Actually, the statistical
interpretation of quantum mechanics follows strictly Hertz philosophical
attitude of accepting the formulae but feeling free to produce some
argument that makes them acceptable. We have called this ``the student's
attitude'' \citep{sola24-perdida} since it replicates the attitude
of the student that is forced to accept the course perspective and
has to find a form to legitimise the mandate of the authority: the
result is given and the justification (at the level of plausibility)
has to be provided. 

The standard QM then has to find a meaning for its formulae and such
task has been called ``interpretation of QM''. All the interpretations
that we know about, rest on the idea of point particles. An idea that
allows for the immediate creation of classical images that suggest
new formulae such as those entering in the atomic spectra of the Hydrogen
atom. Since the accelerated electron (and proton) cannot radiate in
QM by ukase\footnote{ukase: edict of the tsar} the radiation rule
has to be set as a separated rule. The atom, according to the standard
theory, can then be in an ``mixed energy'' state until it is measured
(meaning to put it in contact with the Laboratory) which makes the
state represented by the wave-function to collapse (in the simplest
versions). While there are more sophisticated\footnote{We imply the double meaning as elaborated and false. The first as
current use and the second implied in the etymological root ``sophism''
in the ancient Greek: \textgreek{σοφιστικός} (sophistikós),
latin: sofisticus.}interpretations, none of them incorporates the measurement process
to QM. It remains outside QM.

The idea of isolating an atom deserves examination. One can regard
as possible to set up a region of space in which EM influences from
outside the region can be compensated to produce an effective zero
influence. In the simplest form we think of a Faraday cage and perhaps
similar cages shielding magnetic fields. In the limit, idealised following
Galileo, we have an atom isolated from external influences. But this
procedure is only one half of what is needed. We need the atom to
be unable to influence the laboratory, including the Faraday cage.
And we cannot do anything onto the atom because so doing would imply
it is not isolated. This is, we must rest upon the voluntary cooperation
of the atom to have it isolated. The notion of an isolated atom is
thus shown to be a fantasy. If the atom, as the EM system that it
is, is in the condition of producing radiation, it will produce it
whether the Faraday cage is in place or not. The condition imposed
by the ukase is metaphysical (i.e., not physical, nor the limit of
physical situations). Actually, if such condition is imposed to Hamilton's
principle, the Schrödinger's time-evolution equation is recovered
to the price that the consilience with Electromagnetism and Classical
Mechanics is lost.

It was Schrödinger who dug more deeply into the epistemological problems
of QM. His cat, now routinely killed in every QM course/book was a
form of ridiculing the, socially accepted, statistical interpretation
of QM: 
\begin{quotation}
One can even set up quite ridiculous cases. A cat is penned up in
a steel chamber, along with the following device (which must be secured
against direct interference by the cat): in a Geiger counter there
is a tiny bit of radioactive substance, so small, that perhaps in
the course of the hour one of the atoms decays, but also, with equal
probability, perhaps none; if it happens, the counter tube discharges
and through a relay releases a hammer which shatters a small flask
of hydrocyanic acid. If one has left this entire system to itself
for an hour, one would say that the cat still lives if meanwhile no
atom has decayed. The psi-function of the entire system would express
this by having in it the living and dead cat (pardon the expression)
mixed or smeared out in equal parts.\citep{schrodinger1980present}
\end{quotation}
Schrödinger understood as well that the need for a statistical interpretation
was rooted in the assumption of point particles \citep{schr95}.

In the present work, the conservation of energy of a quantum particle
is described in the form in which Faraday conceived matter as the
inferred part of the matter-action pair and it leads directly to the
transition rule: ${\cal E}_{I}^{f}+{\cal E}_{EM}^{f}={\cal E}_{I}^{i}+{\cal E}_{EM}^{i}$
with $i,f$ meaning initial and final states while $I,EM$ mean internal
and EM energies. In the early times of QM, Bohr showed that the Balmer
and Rydberg lines of the Hydrogen atom corresponded to this rule.
This is to say that the observed values summarised in Rydberg's formula
agree in value with the equation, yet the equation was not part of
the theory because the ``emission'' of light was conceived as the
result of external influences since the isolated atom could not decay
and in such way it would not collapse with the electron falling onto
the proton.

\subsection*{Cognitive surpass as a constructive concept}

The concept of \emph{cognitive surpass} \citep{piag89} is a guiding
idea of the present work. It is a step in the development of our cognition
by which what was previously viewed as diverse is recognised as a
particular occurrence of a more general, abstract form. This cognitive
step is dialectic in nature and displays the dialectic relation between
particular and general. Cognitive surpass involves abstraction and/or
generalisation, it does not suppress the preexisting theories but
rather incorporates them as particular instances of a new, more general
theory. It works in the direction of the unity of science\citep{cat24-unity}
and as such, its motion opposes the specialism characteristic of the
instrumentalist epistemology. Cognitive surpass is a key ingredient
of the abductive process since it is \emph{ampliative} (i.e., it enlarges
the cognitive basis) as opposed to the introduction of ad-hoc hypotheses
to compensate for unexplained observations. 

We notice that current physics hopes for a unified theory but little
to no progress has been made in such direction. This matter should
not come as a surprise since the requirements for understanding were
downgraded to the level of analogy when adopting instrumentalism (see
\citep{boltzmann1974philosphical,mach12-popular}). Analogies hint
us for surpass opportunities connecting facts at the same level, without
producing the surpass. For example, Quantum Electrodynamics superimposes
quantification to electromagnetism making mechanical analogies and
populating an immaterial ether, the space, with immaterial analogous
of classical entities (e.g., \emph{springs}). What is preserved in
this case is the analogy with classical material bodies, the recourse
to imagination/fantasy.

\subsection*{Plan of this work}

In Part I of this work we start from the initial insight by de Broglie
regarding the relation of quantum mechanics and electromagnetism as
Schrödinger did, yet we soon depart from de Broglie as he believed
that Special Relativity was the keystone of electromagnetism. Rather,
we explore the connection with relational electromagnetism. We also
depart from Schrödinger's developments since we will not explore mechanical
analogies neither in meaning nor in formulae. Actually, we show that
de Broglie's starting point leads directly to a rewriting of Electromagnetism
producing a cognitive surpass of the Lorenz-Lorentz variational formulation
\citep[Theorem 3.2,][]{solari2022symmetries} which is behind Maxwell's
equations as propagation in the vacuum of fields with sources (intensification)
in the matter. The same variational principle produces the Lorentz
Force. We show that the variational principle that produces Schrödinger's
evolution equation is the complementary part of one, unifying, variational
principle. The synthesis produced corresponds to the unification of
the quantum entity in the terms proposed by Faraday (action-matter
duality) being the entity not a mathematical point but rather extended
in space in a small intensification region and reaching as far as
its action reaches. We further show that the classical motion under
a Lorentz force corresponds to particular variations of the action.
The classical dynamical equation for the angular momentum is obtained
by the same methods, as well as the conservation of charge and the
total energy (that of the localised entity and its associated electromagnetic
field). It is shown as well that non-radiative states correspond to
those of Schrödinger's equation, matching Bohr's primitive intuition.
In this construction there is no room for free interpretations; measurements
correspond to interactions between the entity and the laboratory described
according to the laws of electromagnetism, hence the measurement process
is completely within the formulation and requires no further postulates. 

In Part II we show how this conception is extended to a proton and
electron interacting to form an Hydrogen atom whose energy levels
calculated by a systematic computation of interactions, require no
patching elements such as gyromagnetic factors or ``relativistic
corrections''. 

The QM we construct allows for a better understanding of the so-called
spin-orbit interaction, which in standard QM couples the spin and
the ``orbit'' of the electron, being then a ``self-interaction''
(an oxymoron) of the electron. We show that when properly and systematically
considered it represents a coupling between the proton and the electron
as all other couplings considered.

\part{The theory}

\section{Abduction of quantum (wave) mechanics \label{subsec:Abduction-of-quantum}}

We begin by highlighting how QM enters the scene as a requirement
of consistence with electromagnetism. Consider a point-charge, $q,$
moving at constant velocity $v$ with respect to a system that acts
as a reference. We associate with this charge the relational charge-current
density
\[
\left(\begin{array}{c}
Cq\\
qv
\end{array}\right)\delta(x-vt)=\left(\begin{array}{c}
C\rho\\
j
\end{array}\right)(x,t).
\]
The association of current with charge-times-velocity goes back to
Lorentz, Weber and possibly further back in time. 

The velocity can then be written as the quotient of the amplitude
of the current divided the charge, ${\displaystyle \beta=\frac{qv}{Cq}}$.
At the same time, Lagrangian mechanics associates the velocity with
the momentum, $p$, of a free moving body by a relation of the form,
${\displaystyle \beta=\frac{p}{Cm}}$ . We propose that charge and
mass are constitutive attributes of microscopic systems and hence
\[
\left(\begin{array}{c}
mC\\
p
\end{array}\right)=\frac{m}{q}\left(\begin{array}{c}
qC\\
qv
\end{array}\right)
\]
will behave as the charge-current density when intervening in electromagnetic
relations. \footnote{A sort of myth emerges at this point. Lorentz \citep{lorentz1904electromagnetic}
tried to fix the electromagnetism conceived by consideration of the
ether. After lengthy mathematical manipulations of Maxwell's equations
he arrived to the point in which the force exerted in the direction
of motion held a different relation with the change of mechanical
momentum than the force exerted in the perpendicular direction. He
then substituted a constant velocity introduced as a change of reference
frame with the time-dependent velocity of the electron with respect
to the ether. This is a classical utilitarian approach, the formula
is applied outside its scope although the mathematics that led to
it does not hold true if the velocity is not constant. This is the
origin of Lorentz' mass formula as applied for example to the Kauffman's
experiment. The same utilitarian operation is used to produce the
relativistic mass formulae. However, if this new mass were to be used
in the derivation of Lorentz' force \citep{lorentz1892CorpsMouvants},
the usual expression of the force would not hold. The story illustrates
well the ``advantages'' of the utilitarian approach: it frees the
scientist of requirements of consistency. In contrast, relational
electromagnetism acknowledges that when in motion relative to the
source we perceive EM fields still as EM fields but with characteristic
changes as it is made evident by the Doppler effect. This allows to
propose the form of the perceived fields when in motion at a constant
relative velocity with respect to the source \citep{solari2022symmetries}.}

If rather than a point particle we accept an extended body moving
at constant velocity without deforming and represented by a charge-current
density
\[
\left(\begin{array}{c}
Cq\\
qv
\end{array}\right)\rho(x,t)
\]
where $\int d^{3}x\,\rho(x,t)=1$, we find that the conditions posed
above can be satisfied whenever
\[
\rho(x,t)\propto\int d^{3}k\,\phi(k)\exp(i(k\cdot x-wt))
\]
with $w=f(k)$. If we further associate the frequency with the (kinetic)
energy in the form $E=\hbar w$, as de Broglie did \citep{de1924recherches},
and write $p=\hbar k$ we have 
\begin{eqnarray*}
E & = & \hbar f(\frac{p}{\hbar}).
\end{eqnarray*}

Since we expect $\rho(x,t)$ to be real, it is clear that this association
is not enough to rule out the unobservable negative energies, since\\
 ${\displaystyle \rho=\rho^{*}\Leftrightarrow\phi(k)\exp(i(k\cdot x-\frac{E}{\hbar}t))=\phi^{*}(k)\exp(i((-k)\cdot x-(-\frac{E}{\hbar})t))}$.
Negative energies would be required and they would rest on an equal
footing as positive ones. Introducing the quantum-mechanical wave-function
$\psi$ satisfying 
\begin{eqnarray}
\psi(x,t) & = & \left(\frac{1}{2\pi}\right)^{3/2}\int d^{3}k\,\Phi^{*}(k)\exp(\frac{i}{\hbar}(p\cdot x-Et))\nonumber \\
 & = & \left(\frac{1}{2\pi\hbar}\right)^{3/2}\int d^{3}p\,\Phi^{*}(\frac{p}{\hbar})\exp(\frac{i}{\hbar}(p\cdot x-Et))\label{eq:wavefunction}\\
\rho(x,t) & = & q\ \psi(x,t)^{*}\psi(x,t)\nonumber 
\end{eqnarray}
we automatically obtain a real charge-density without the imposition
of negative energies (where for free particles $E\equiv T$, the kinetic
energy, whose expression will be derived below). 

We retain from traditional Quantum Mechanics the idea of expressing
physical observables as real valued quantities  associated to an \emph{operator}
integrated over the whole entity. 

We recall that the quantum entity is a unity and \emph{not} a mathematical
point. Hence, objects like $\left(\psi^{*}\left(-i\hbar\nabla\psi\right)\right)$
do not necessarily have physical (measurable) meaning. Further, observables
depending only on $x$ are multiplicative operators in front of $\psi(x,t)$.
For the case of the position operator $x$ we have \footnote{Integrals such as ${\displaystyle \langle\psi^{*}x\psi\rangle=\int d^{3}x\left(\psi^{*}x\psi\right)}$
or ${\displaystyle \langle\psi^{*}(-i\hbar\nabla\psi)\rangle=\int d^{3}x\left(\psi^{*}(-i\hbar\nabla\psi)\right)}$
have been called \emph{expectation values} in the traditional probability
interpretation of Quantum Mechanics.}, {\footnotesize{}
\begin{eqnarray*}
\langle\hat{x}\rangle & = & \int d^{3}x\ \psi^{*}(x,t)\ x\ \psi(x,t)\\
 & = & \int d^{3}x\ d^{3}p\ d^{3}p^{\prime}\ \frac{1}{\left(2\pi\hbar\right)^{3}}\Phi^{*}(p^{\prime})\exp(\frac{i}{\hbar}(p^{\prime}\cdot x-Et))\ x\ \Phi(p)\exp(-\frac{i}{\hbar}(p\cdot x-Et))\\
 & = & \int d^{3}p\ d^{3}p^{\prime}\ \Phi^{*}(p^{\prime})\ \left(i\hbar\nabla_{p}\ \Phi(p)\right)\int d^{3}x\ \frac{1}{\left(2\pi\hbar\right)^{3}}\exp(\frac{i}{\hbar}(p-p^{\prime})\cdot x)\\
 & = & \int d^{3}p\ \Phi^{*}(p)\ (i\hbar\nabla_{p}\Phi(p))
\end{eqnarray*}
} while\footnote{Throughout this work the notations $\partial_{t}\psi$, $\dot{\psi}$
and $\psi_{,t}$ all denote the same concept, namely partial derivative
with respect to time.} 
\begin{eqnarray*}
\hat{E} & = & i\hbar\partial_{t}.
\end{eqnarray*}
De Broglie's proposal and eq.\eqref{eq:wavefunction} relate $x$
and $p$ through a Fourier transform and hence the momentum operator
becomes $\hat{p}=m\hat{v}=-i\hbar\nabla$.  Thus, in the case of a
free charged particle we arrive to Schrödinger's equation 
\begin{equation}
i\hbar\partial_{t}\psi=\hat{E}\psi.\label{eq:free-electron}
\end{equation}

We have yet to find $f(k)$. If we consider that eq.\eqref{eq:free-electron}
is the definition of the Hamiltonian operator $\hat{H}\equiv\hat{E}$,
and recall that the operator associated to the current density is
${\displaystyle q\hat{v}=\frac{q}{m}\hat{p}}$. Hence, the conservation
of charge, also known as the \emph{continuity equation}, $\partial_{t}\rho+\nabla\cdot j=0$,
produces the relations 
\begin{eqnarray*}
q\partial_{t}\rho & = & q\left[\psi^{*}\partial_{t}\psi+\partial_{t}\psi^{*}\psi\right]=\frac{q}{i\hbar}\left[\psi^{*}(\hat{H}\psi)-(\hat{H}\psi)^{*}\psi\right]\\
 & = & -\nabla\cdot\frac{1}{2}\frac{q}{m}\left[\psi^{*}\hat{p}\psi+(\hat{p}\psi)^{*}\psi\right]
\end{eqnarray*}
and expanding the r.h.s. we can see that the only solution is:
\begin{eqnarray*}
\hat{p} & = & -i\hbar\nabla\\
\hat{H} & = & \frac{\hat{p}^{2}}{2m}+U(x)=-\frac{\hbar^{2}}{2m}\Delta+U(x)\\
j & = & \frac{q}{2m}\left(\psi^{*}\left(-i\hbar\nabla\psi\right)+\left(-i\hbar\nabla\psi\right){}^{*}\psi\right)=-\frac{i\hbar q}{2m}\left(\psi^{*}\left(\nabla\psi\right)-\left(\nabla\psi\right){}^{*}\psi\right)
\end{eqnarray*}
where $U(x)$ is any scalar function of the position. We adopt $U(x)=0$
as for a free-particle there is no justification for a potential energy.
We have then found the free-particle Hamiltonian consistent with de
Broglie's approach and the assumption $\rho(x,t)=|\psi(x,t)|^{2}$.\footnote{In his original work, de Broglie \citep{de1924recherches} assumed
the relation, $E^{2}=(m^{2}C^{2}+p^{2})C^{2}$ proposed by special
relativity on an utilitarian basis as already commented. It was later
Schrödinger \citep{schrodinger1926undulatory} who proposed the classical
form based on the correspondence with Newton's mechanics.}

In summary, QM results from requirements of compatibility with electromagnetism
following an abduction process in which the hypothesis $E=\hbar w$
results as an extension of the photoelectric relation. As a consequence,
any attempt to superimpose further formal relations with electromagnetism
such as requiring invariance or equivariance of forms under the Poincaré-Lorentz
group is bound to introduce more problems than solutions.

\subsection{Expression of the electromagnetic potentials\label{subsec:Potentials}}

\subsubsection{Physical Background}

In this section our goal is to develop an electromagnetic theory of
microscopic systems encompassing both QM and Maxwell's electrodynamics.
Hence, we will compute Maxwell's EM interaction energy, either between
the microscopic system and external EM-fields or between different
fields arising within the system. We display first some classical
results that are relevant for the coming computations.

Maxwell's electromagnetic \emph{interaction energy} between entities
$1$ and $2$ reads:

\textbf{
\begin{equation}
{\cal E}=\int d^{3}x\left[\epsilon_{0}E_{1}\cdot E_{2}+\frac{1}{\mu_{0}}B_{1}\cdot B_{2}\right].\label{eq:EM-energy}
\end{equation}
}The classical EM potentials originating in a charge density $\left|\psi\right|^{2}$
and a current $-\frac{i\hbar q}{2m}\left(\psi^{*}(\nabla_{y}\psi)-(\nabla_{y}\psi)\psi\right)(y,t)$,
as defined in the Lorenz gauge, read
\begin{eqnarray*}
V(x,t) & = & \frac{q}{4\pi\epsilon_{0}}\int\!d^{3}y\ \frac{1}{|x-y|}\left|\psi(y,t-|x-y|/C)\right|^{2}
\end{eqnarray*}
 for the scalar potential (where as usual $C^{2}=\left(\mu_{0}\epsilon_{0}\right)^{-1}$),
while the vector potential is: 
\begin{eqnarray*}
A(x,t) & =-{\displaystyle \frac{i\hbar}{2}} & \frac{\mu_{0}q}{4\pi m}\int\!d^{3}y\ \frac{1}{|x-y|}\left(\psi^{*}(\nabla_{y}\psi)-(\nabla_{y}\psi)\psi\right)(y,t-|x-y|/C).
\end{eqnarray*}
Finally, electric and magnetic fields are obtained through Maxwell's
equations ${\displaystyle E=-\nabla V-\frac{\partial A}{\partial t}}$
and $B=\nabla\times A$. We recall from Maxwell \citep[{see e.g. [630]}][ ]{maxw73}
that all electromagnetic quantities and the wave-function are assumed
to vanish sufficiently fast at infinity. This allows for the use of
partial integrations and Gauss' theorem whenever appropriate.

\subsubsection{Action integral and the least action principle}

Lorentz' electromagnetic action \citep{lorentz1892CorpsMouvants}
is 
\begin{eqnarray}
{\cal A}_{EM} & = & \int_{t_{1}}^{t_{2}}\left\{ {\cal L}_{EM}\right\} ds\label{eq:action-1}\\
{\cal L}_{EM} & = & \frac{1}{2}\int d^{3}x\left\{ \frac{1}{\mu_{0}}B^{2}-\epsilon_{0}E^{2}\right\} =\frac{1}{2}\int d^{3}x\left\{ A\cdot j-V\cdot\rho\right\} \nonumber 
\end{eqnarray}

Equivalence between the field and potential forms is assured by the
followin
\begin{lem}
\label{lem:action-final}Up to an overall function of time and the
divergence of a function vanishing sufficiently fast at infinity,
the electromagnetic action satisfies
\begin{eqnarray*}
{\cal {\cal A}} & = & \frac{1}{2}\int_{t_{0}}^{t_{1}}dt\int\left(\frac{1}{\mu_{0}}B{}^{2}-\epsilon_{0}E{}^{2}\right)\,d^{3}x=\frac{1}{2}\int_{t_{0}}^{t_{1}}dt\,\int\left(A\cdot j-\rho V\right)\,d^{3}x
\end{eqnarray*}
\end{lem}

See Appendix \ref{sec:Proof-of-variational} for a proof.

Unfortunately, Lorentz worked from his expressions in a mixed form
using mathematics and intuitions generated by the hypothesis regarding
the existence of a material medium permeating everything: the ether.
It is from these mathematical objects that he derives the electromagnetic
force. Since the original derivation is marred by the ether, the force
was adopted as an addition to the collection formed by Maxwell's equations
and the equation of continuity.

Contrasting to Lorentz, Lorenz \citep{lore67} related light to electromagnetic
currents in a body, being the propagation in the form of waves as
he had previously established by experimental methods \citep{lore61}\footnote{Notice that Lorenz ideas were published two years before and two years
after Maxwell seminal paper \citep{maxw64} and at least ten years
before Hertz' famous experiments of 1887-1888. The historical narrative
that reaches us attributing to Maxwell and Hertz this association
is historically incorrect as it is incorrect the often stated opinion
(for example in Lorentz \citep{lorentz1892CorpsMouvants}) that Faraday
supported the idea of the ether when actually he had warned about
the risks of inventing metaphysical objects and has provided the alternative
of his theory of ``vibrating rays'' \citep{fara55}.

Lorenz, Faraday as much as Newton, were ahead of their times and those
that followed recast their learnings in forms apt for the reader to
understand (called ``didactic transposition'' when teaching). A
first example of misreading Faraday is precisely Maxwell \citep[{V. ii, p. 177 [529]}][ ]{maxw73}
who wrote about Faraday's idea: ``He even speaks {*} of the lines
of force belonging to a body as in some sense part of itself, so that
in its action on distant bodies it cannot be said to act where it
is not. This, however, is not a dominant idea with Faraday. I think
he would rather have said that the field of space is full of lines
of force, whose arrangement depends on that of the bodies in the field,
and that the mechanical and electrical action on each body is determined
by the lines which abut on it. {*} \citep[p, 447;][]{fara55}'' .
Goethe has prevented us: ``You don’t have to have seen or experienced
everything for yourself; but if you want to trust another person and
his descriptions, remember that you are now dealing with three factors:
with the matter itself and two subjects.'' \citep[570,][]{goethe32}.}. His idea can be expressed in terms of a variation of the action
\ref{eq:action-1} with respect to the potentials where the potentials
have to match the values in the region of their source. We have called
this result ``the Lorenz-Lorentz theorem'' \citep[Theorem 3.2][]{solari2022symmetries}.
The variation accounts for the propagation of electromagnetic fields
and imposes the condition of conservation of charge as a result ultimately
produced by the observations on the propagation of the electromagnetic
action. The same action \ref{eq:action-1} when varied as Lorentz
proposed with respect to the charge-current reveals the Lorentz force.
The details of the derivation can be found in \citep{natiello2021relational}.

Lorentz's action accounts for the propagation of electromagnetic effects
but not for the sources, in contrast Schrödinger's equation for the
free charge can be derived from a variational principle
\[
{\cal A}_{QM}=\int_{t_{0}}^{t_{1}}ds\left\langle \frac{1}{2}\left(\psi^{*}\left(i\hbar\dot{\psi}\right)+(i\hbar\dot{\psi})^{*}\psi\right)+\frac{\hbar^{2}}{2m}\psi^{*}\Delta\psi\right\rangle 
\]
where $\left\langle \right\rangle $ indicates integration in all
space and $\Delta$ is Laplace's operator. The Schrödinger equation
is recovered when the QM-action is varied with respect to $\psi$
(or $\psi^{*}$): 
\begin{lem}
\label{lem:Quantum}${\cal A}_{QM}$ is stationary with respect to
variations of $\psi^{*}$ such that $\delta\psi^{*}(t_{0})=0=\delta\psi^{*}(t_{1})$
if and only if ${\displaystyle i\hbar\dot{\psi}}=-{\displaystyle \frac{\hbar^{2}}{2m}}\Delta\psi$.
\end{lem}

The hope is then that by adding both actions, ${\cal A}_{EM}$ that
almost ignores the dynamics of the sources and ${\cal A}_{QM}$ that
ignores the associated fields we can propose a synthetic action integral{\small{}
\begin{eqnarray}
{\cal A} & = & {\cal A}_{EM}+{\cal A}_{QM}\nonumber \\
 & = & \int_{t_{0}}^{t_{1}}ds\left\langle \frac{1}{2}\left(\psi^{*}\left(i\hbar\dot{\psi}\right)+(i\hbar\dot{\psi})^{*}\psi\right)+\frac{\hbar^{2}}{2m}\psi^{*}\Delta\psi+\frac{1}{2}\left(\frac{1}{\mu_{0}}B^{2}-\epsilon_{0}E^{2}\right)\right\rangle \label{eq:TheAction}
\end{eqnarray}
}no longer separating effects of sources, since, as stated in subsection
\ref{subsec:Potentials}, in the present construction all of current,
charge-density and electromagnetic potentials depend of the wave-function
$\psi$. Hamilton's variational principle applied to the action integral
${\cal A}$ yields all the dynamical equations in EM, and QM. 

\subsection{Variations of the action\label{subsec:Variations-of-the}}

The variations we are going to consider will be the product of an
infinitesimal real function of time denoted $\epsilon$ when it is
a scalar and $\bar{\epsilon}$ (a vector of arbitrary constant direction,
$\bar{\epsilon}=\epsilon k,\epsilon r,\epsilon\Omega$) when it refers
to a wave number, position or solid-angle respectively; times a function
$\mu$ often of the form $\mu=\hat{O}\psi$ for some appropriate operator
$\hat{O}$. The variation of the wave-function is reflected directly
in variations on the charge-density, current, and fields. We will
use for them the symbols $\delta\rho,\delta j,\delta E,\delta B$
for a set of variations that satisfy Maxwell's equations.

Hence, according to the previous paragraph, the variation $\delta\psi=\epsilon\mu$
is reflected in a variation of the EM quantities as follows:
\begin{enumerate}
\item Variation of charge-density: $\delta\rho=\epsilon q\left(\mu^{*}\psi+\text{\ensuremath{\psi^{*}\mu}}\right)\equiv\epsilon\ts{X}$.
\item Variation of current: $\delta j=\epsilon\ts{J}_{1}+\dot{\epsilon}\ts{J}_{2}$.
We will discuss in due time the connection between $\delta\psi$ and
$\delta j$.
\end{enumerate}
The continuity equation then reads
\[
\left(\delta\rho\right)_{,t}+\nabla\cdot\delta j=0,
\]
which discloses into 
\begin{eqnarray*}
\ts{X}+\nabla\cdot\ts{J}_{2} & = & 0\\
\ts{X}_{,t}+\nabla\cdot\ts{J}_{1} & = & 0
\end{eqnarray*}
thus demanding $\nabla\cdot\ts{J}_{1}-\nabla\cdot\ts{J}_{2,t}=0$
and suggesting $\ts{J}_{2}=-\epsilon_{0}\ts{E}$, with $\epsilon_{0}\nabla\cdot\ts{E}=\ts{X}$
for some electrical field that needs yet further specification. 

We develop here the mathematical results of this manuscript. Proofs
can be found in Appendix \ref{sec:Proof-of-variational}.
\begin{lem}
\label{lem:Prep}(Preparatory lemma). Under the previous definitions
and the conditions $\ts{X}+\nabla\cdot\ts{J}_{2}=0$ and $\left(\ts{X}\right)_{,t}+\nabla\cdot\ts{J}_{_{1}}=0$,
complying with the continuity equation, a variation $\delta\psi=\epsilon\mu$
of the wave-function produces the varied action integral{\small{}
\begin{eqnarray*}
\delta{\cal A} & = & \delta\int_{t_{0}}^{t_{1}}ds\left\{ \left\langle \frac{1}{2}\left(\psi^{*}\left(i\hbar\dot{\psi}\right)+(i\hbar\dot{\psi})^{*}\psi\right)+\frac{\hbar^{2}}{2m}\psi^{*}\Delta\psi\right\rangle +\frac{1}{2}\left\langle \frac{1}{\mu_{0}}B^{2}-\epsilon_{0}E^{2}\right\rangle \right\} \\
 & = & \int_{t_{0}}^{t_{1}}\epsilon\ ds\left\langle \left(\mu^{*}\left(i\hbar\dot{\psi}\right)+(i\hbar\dot{\psi})^{*}\mu\right)+\frac{\hbar^{2}}{2m}\left(\mu^{*}\Delta\psi+\mu\Delta\psi^{*}\right)\right\rangle +\\
 &  & +\int_{t_{0}}^{t_{1}}\epsilon\ ds\left\langle A\cdot\ts{J}_{1}-\left(A\cdot\ts{J}_{2}\right)_{,t}-V\ts{X}\right\rangle ,
\end{eqnarray*}
}where the fields are denoted as \textbf{$B=B_{m}+B_{r}$} and $E=E_{m}+E_{r}$,
with the index $m$ corresponding to the microscopic system to be
varied and $r$ denotes the ``outside world''. $A,V$ are the electromagnetic
potentials. 
\end{lem}

\begin{coro}
\label{cor:J2=00003DE}Without loss of generality, we set $\ts{J}_{2}=-\epsilon_{0}\ts{E}$,
where $\epsilon_{0}\nabla\cdot\ts{E}=\ts{X}$.
\end{coro}

We will address different relations in a few related results which
are stated separately because of expository convenience. The details
of the proofs will be displayed in Appendix \ref{sec:Proof-of-variational}. 
\begin{thm}
\label{thm:Classical}(Classical variations theorem). Under the assumptions
of Lemma \ref{lem:Prep}, the null variation of the action
\[
{\cal A}=\int_{t_{1}}^{t_{2}}ds\left(\left\langle \frac{1}{2}\left(\psi^{*}\left(i\hbar\dot{\psi}\right)+(i\hbar\dot{\psi})^{*}\psi\right)+\frac{\hbar^{2}}{2m}\psi^{*}\Delta\psi\right\rangle +\frac{1}{2}\left\langle \frac{1}{\mu_{0}}B^{2}-\epsilon_{0}E^{2}\right\rangle \right)
\]
as a consequence of a variation of the wave-function as: 
\[
\delta\psi=\epsilon\psi_{,t},\,\left(i\epsilon k\cdot x\right)\psi,\,\left(\epsilon\bar{x}\cdot\nabla\psi\right),\,\left(\epsilon\Omega\cdot\left(x\times\nabla\psi\right)\right)
\]
 --as summarised in Table \ref{tab:Variations}-- results in the
corresponding equations of motion displayed in Table \ref{tab:Variation-results}.

\begin{table}
\begin{tabular}{|c|c|c|c|c|}
\hline 
{\scriptsize{}Variation} & {\footnotesize{}$\delta\psi$} & {\footnotesize{}$\ts{X}$} & {\footnotesize{}$\ts{J}_{1}$} & {\footnotesize{}$\ts{J}_{2}$}\tabularnewline
\hline 
\hline 
{\scriptsize{}time} & {\footnotesize{}$\epsilon\partial_{t}\psi$} & {\footnotesize{}$\rho_{,t}$} & {\footnotesize{}$-2\epsilon_{0}\ts{E}{}_{,t}-\ts{j}=j_{,t}$} & {\footnotesize{}$j$}\tabularnewline
\hline 
{\scriptsize{}position} & {\footnotesize{}$\left(i\epsilon k\cdot x\right)\psi$} & {\footnotesize{}$0$} & {\footnotesize{}$0$} & {\footnotesize{}$0$}\tabularnewline
\hline 
{\scriptsize{}displacement} & {\footnotesize{}$\epsilon r\cdot\nabla\psi$} & {\footnotesize{}$r\cdot\nabla\rho$} & {\footnotesize{}$\left(r\cdot\nabla\right)j=\ts{j}$ } & {\footnotesize{}$-r\rho$}\tabularnewline
\hline 
{\scriptsize{}rotation} & {\footnotesize{}$\epsilon\Omega\cdot\left(x\times\nabla\psi\right)$} & {\footnotesize{}$\Omega\cdot\left(x\times\nabla\rho\right)$} & {\footnotesize{}$\left(\Omega\cdot\left(x\times\nabla\right)\right)j-\Omega\times j=\ts{j}$} & {\footnotesize{}$-\Omega\times\left(x\rho\right)$}\tabularnewline
\hline 
\end{tabular}

\caption{\label{tab:Variations}Variations}
\end{table}

\begin{table}
\begin{tabular}{|c|c|c|}
\hline 
{\scriptsize{}Variation/result} & {\footnotesize{}$\delta\psi$} & {\scriptsize{}Equation of motion}\tabularnewline
\hline 
\hline 
{\scriptsize{}time/energy} & {\footnotesize{}$\epsilon\partial_{t}\psi$} & {\footnotesize{}$\partial_{t}\left\langle -\frac{\hbar^{2}}{2m}\psi^{*}\Delta\psi+\frac{1}{2}\left(\epsilon_{0}E^{2}+\frac{1}{\mu_{0}}B^{2}\right)\right\rangle =0$}\tabularnewline
\hline 
{\scriptsize{}position/velocity} & {\footnotesize{}$\left(i\epsilon k\cdot x\right)\psi$} & {\footnotesize{}${\displaystyle \partial_{t}\left\langle \psi^{*}x\psi\right\rangle =\frac{\left\langle j\right\rangle }{q}}$}\tabularnewline
\hline 
{\scriptsize{}displacement/momentum} & {\footnotesize{}$\epsilon r\cdot\nabla\psi$} & {\footnotesize{}${\displaystyle \frac{m}{q}}\partial_{t}\left\langle j\right\rangle =\left\langle j\times B+\rho E\right\rangle $}\tabularnewline
\hline 
{\scriptsize{}rotation/angular momentum} & {\footnotesize{}$\epsilon\Omega\cdot\left(x\times\nabla\psi\right)$} & {\footnotesize{}${\displaystyle \frac{m}{q}}\partial_{t}\left\langle x\times j\right\rangle =\left\langle x\times\left(\rho E+j\times B\right)\right\rangle $}\tabularnewline
\hline 
\end{tabular}

\caption{\label{tab:Variation-results}Equations of motion.}
\end{table}
\end{thm}

The entries in table \ref{tab:Variations} were produced following
intuitions based upon the notion of conjugated variables in classical
mechanics. A displacement in time is associated with the law of conservation
of energy, one in space with the law of variation of linear momentum
and an angular displacement with angular momentum. The variation leading
to the derivative of the position is of an abductive nature since
what it is really known is the cinematic relation. Hence, the variation
is obtained as the answer to the question: which variation produces
the cinematic relation $\left\langle x\right\rangle _{,t}=\left\langle j\right\rangle /q$? 

The proposed variations do not follow a general pattern. Hence, we
expect that there might be alternative forms of variations that lead
to identical results which in addition present consilience. Moreover,
we expect that the energy would take the place of the Lagrangian since
this is one of the intuitions based upon experimental observations
that lead to quantum mechanics. We present the result of this synthetic
steps in the next theorem. 

There are four matters that have to be put together to complete the
construction of Quantum Mechanics:
\begin{enumerate}
\item Changing the Lagrangian presentation into a Hamiltonian one.
\item Showing that the Hamiltonian presentation has consilience and produces
the particular results of Theorem \ref{thm:Classical} as well (it
is a generalisation/correction of the particular cases shown).
\item Showing that this setting produces the usual eigenvalue-eigenvector
equations of Quantum Mechanics for the stationary wave-functions.
\item Making more explicit the relation between the variation of the wave-function
and the variation of the electromagnetic current.
\end{enumerate}
We state these results as three Lemmas and a Corollary.
\begin{lem}
(Hamiltonian form)\label{lem:Hamiltonian-form.} Let the set of functions
$\ts{X},\ts{j},\ts{E},\ts{B}$ satisfy Maxwell equations with $\delta\rho=\epsilon\ts{X}=\epsilon q\left(\mu^{*}\psi+\psi^{*}\mu\right)$
as in Lemma \ref{lem:Prep} . If we let the current be $\delta j=\epsilon\ts{J}_{1}+\dot{\epsilon}\ts{J}_{2}$,
with
\begin{eqnarray*}
\ts{J}_{2} & = & -\epsilon_{0}\ts{E}\\
\ts{J}_{1} & = & -2\epsilon_{0}\ts{E}_{,t}-\ts{j}
\end{eqnarray*}
the null variation of the action integral satisfies
\begin{equation}
\left\langle \mu^{*}\left(i\hbar\dot{\psi}\right)+\left(i\hbar\dot{\psi}\right)^{*}\mu\right\rangle +\frac{\hbar^{2}}{2m}\left\langle \mu^{*}\Delta\psi+\mu\Delta\psi^{*}\right\rangle -\left\langle \frac{1}{\mu_{0}}B\cdot\ts{B}+\epsilon_{0}E\cdot\ts{E}\right\rangle =0,\label{eq:Hamilton}
\end{equation}
where ${\displaystyle -\frac{\hbar^{2}}{2m}\left\langle \mu^{*}\Delta\psi+\mu\Delta\psi^{*}\right\rangle +\left\langle \frac{1}{\mu_{0}}B\cdot\ts{B}+\epsilon_{0}E\cdot\ts{E}\right\rangle }\equiv{\cal H}$
is recognised as the Hamiltonian of the problem.
\end{lem}

\begin{coro}
\label{cor:hamiltonian}The form of the variations appropriated to
the Hamiltonian form corresponding to the results of Theorem \ref{thm:Classical}
(Table \ref{tab:Variation-results}) is the one given in Table \ref{tab:hamiltonian-variations}.
\begin{table}
\begin{tabular}{|c|c|c|c|c|c|}
\hline 
{\scriptsize{}Variation} & {\scriptsize{}$\delta\psi$} & {\scriptsize{}$\ts{X}$} & {\scriptsize{}$\ts{j}$} & {\scriptsize{}$\epsilon_{0}\ts{E}$} & {\scriptsize{}$\nabla\times\ts{B}$}\tabularnewline
\hline 
\hline 
{\scriptsize{}time} & {\scriptsize{}$\epsilon\partial_{t}\psi$} & {\scriptsize{}$\rho_{,t}$} & {\scriptsize{}$j_{,t}$} & {\scriptsize{}$E_{,t}$} & {\scriptsize{}$\text{\ensuremath{\nabla\times B_{,t}}}$}\tabularnewline
\hline 
{\scriptsize{}position} & {\scriptsize{}$\left(i\epsilon k\cdot x\right)\psi$} & {\scriptsize{}$0$} & {\scriptsize{}$0$} & {\scriptsize{}$0$} & {\scriptsize{}$0$}\tabularnewline
\hline 
{\scriptsize{}displacement} & {\scriptsize{}$\epsilon r\cdot\nabla\psi$} & {\scriptsize{}$\left(r\cdot\nabla\right)\rho$} & {\scriptsize{}$\left(r\cdot\nabla\right)j$} & {\scriptsize{}$r\rho$ } & {\scriptsize{}$-\nabla\times\left(r\times j\right)$}\tabularnewline
\hline 
{\scriptsize{}rotation} & {\scriptsize{}$\epsilon\Omega\cdot\left(x\times\nabla\psi\right)$} & {\scriptsize{}$\Omega\cdot\left(x\times\nabla\rho\right)$} & {\scriptsize{}$\left(\Omega\cdot\left(x\times\nabla\right)\right)j-\Omega\times j$} & {\scriptsize{}$\Omega\times\left(x\rho\right)$} & {\scriptsize{}$-\nabla\times\left(\left(\Omega\times x\right)\times j\right)$}\tabularnewline
\hline 
\end{tabular}

\caption{\label{tab:hamiltonian-variations}Variations in Hamiltonian form}
\end{table}
\end{coro}

\begin{lem}
(Variation of a stationary state)\label{lem:stationary}. In the special
case when $\psi$ corresponds to a stationary state, $\psi(x,t)=\exp\left(-i\omega t\right)\psi(x,0)\equiv\exp\left(-i\omega t\right)\psi_{\omega}$
and assuming that the external electromagnetic fields are time-independent,
and $B$ is homogeneous, the variation in Lemma \ref{lem:Hamiltonian-form.}
yields
\begin{equation}
\left\langle \mu^{*}\left(\hbar\omega\psi_{\omega}\right)\right\rangle +\frac{\hbar^{2}}{2m}\left\langle \mu^{*}\Delta\psi_{\omega}\right\rangle -\left\langle \mu^{*}\hat{H}_{EM}\psi_{\omega}\right\rangle =0\label{eq:eigenvalue}
\end{equation}

where 
\[
\hat{H}_{EM}\psi=V\psi-\left(\frac{q}{2m}\right)\left(B\cdot\left(x\times\left(-i\hbar\nabla\right)\right)\psi\right)
\]

Eq.\eqref{eq:eigenvalue} gives the Schrödinger equation for stationary
states. 
\end{lem}

The original expression by Schrödinger was developed setting $B=0$
and assuming that the electron is the quantum entity while the proton
is an ``external'' point charge, responsible for the electrostatic
potential $V$.

In Theorem \ref{thm:Classical} as well as in Corollary \ref{cor:hamiltonian}
we rested on phenomenological ideas (synthetic judgements) to produce
the current associated to the variation. To find a general form for
a set of intuitions appears as almost impossible. We proposed in the
case of the rotations $\ts{j}=\left(\Omega\cdot\left(x\times\nabla\right)\right)j-\Omega\times j$
because the rotation of vectors demands to change the spatial argument
as well as the orientation of the components. The confirmation of
the abductive hypothesis came from the result, which has an intuitively
appealing form. On the contrary, had we forgotten the term $-\Omega\times j$
we would have ended up with a result with no consilience with that
of classical mechanics. We stress that these are rational arguments
which are not of a deductive nature. Nevertheless, there are some
mathematical relations that have to be satisfied since the continuity
equation must be enforced.
\begin{lem}
\label{lem:variations}The variations $\hat{O}\psi$ (with $\hat{O}_{,t}=0$)
that may produce a varied current $\ts{j}$ fulfilling the conditions
of Lemma \ref{lem:Prep} extend as real operators $\Theta$ acting
on forms such as $(\rho,\ j,\ \nabla\cdot j)$ that can be regarded
as real-valued functions of $(\psi^{*},\psi)$ as: {\footnotesize{}
\begin{eqnarray*}
\Theta\rho\left(\psi^{*},\psi\right) & \!\!= & \!\!\rho\left(\left(\hat{O}\psi\right)^{*},\psi\right)+\rho\left(\psi^{*},\left(\hat{O}\psi\right)\right)\\
\Theta j\left(\psi^{*},\psi,\nabla\psi^{*},\nabla\psi\right) & \!\!= & \!\!j\left(\left(\hat{O}\psi\right)^{*},\psi,\left(\hat{O}\nabla\psi\right)^{*},\nabla\psi\right)\!+\!j\left(\psi^{*},\hat{O}\psi,\nabla\psi^{*},\left(\hat{O}\nabla\psi\right)\right)\\
\Theta\left(\nabla\cdot j\left(\psi^{*},\psi,\Delta\psi^{*},\Delta\psi\right)\right) & \!\!= & \!\!j\left(\left(\hat{O}\psi\right)^{*},\psi,\left(\hat{O}\Delta\psi\right)^{*},\Delta\psi\right)\!+\!j\left(\psi^{*},\left(\hat{O}\psi\right),\Delta\psi^{*},\left(\hat{O}\Delta\psi\right)\right).
\end{eqnarray*}
}Then, $\Theta$ preserves the continuity equation as $\Theta\left(\rho_{,t}+\nabla\cdot j\right)=0$
and\linebreak{}
 $\Theta(\nabla\cdot j)=\nabla\cdot\ts{j}$ when the variation satisfies
the relation 
\[
\psi^{*}\left(\left[\Delta,\hat{O}\right]\psi\right)=\psi^{*}\left(\left[\Delta,\hat{O}\right]^{\dagger}\psi\right)+\nabla\cdot K
\]
(where $\dagger$ indicates Hermitian conjugation) for some vector
$K$ satisfying\linebreak{}
 $\int d^{3}x\,\left(\nabla\cdot K\right)=\int_{\Sigma}K\cdot\widehat{d\Sigma}=0$.
In such a case, $\ts{j}=\Theta j+K$. The operator $\hat{O}$ is then
antiHermitian.
\end{lem}

As examples for the application of the previous lemma we use the cases
listed in table \ref{tab:hamiltonian-variations}. The operators associated
to energy, velocity and rotation are $\hat{O}=\left(\partial_{t},(r\cdot\nabla),\Omega\cdot\left(x\times\nabla\right)\right)$.
All of them commute with Laplace's operator, hence $\ts{j}=\Theta j$.
In the case of the position $\Theta j=0,\,\Theta\left(\nabla\cdot j\right)=0$
and the terms including $\left[\nabla,\hat{O}\right]$ add up to zero.

\subsection{Discussion}

In Subsection \ref{subsec:Variations-of-the}we have shown that when
the electromagnetic action outside the locus of masses, charges and
currents as it appears in the Lorenz-Lorentz theorem is complemented
with the action where mass, charge and currents are not necessarily
null, the action proposed encodes Maxwell's electromagnetic laws,
Lorentz' force and Newton's equations of motion, as well as an extension
of Schrödinger equation for stationary states . Thus, electromagnetism
does not question the validity of classical mechanics, nor quantum
mechanics questions the validity of electromagnetism or classical
mechanics. They coexist in a synthetic variational equation.

The difference between Schrodinger's eigenvalue-eigenvector equation
and eq. \eqref{eq:eigenvalue} corresponds to the ``static energy''
which is present in Maxwell's formulation of electromagnetism \citep[{[680] p. 270}][ ]{maxw73}.
The method of construction used by Maxwell is not consistent with
the notion of quanta since it assumes that any charge can be divided
in infinitesimal amounts, the electrostatic energy corresponding to
assembling together the total charge. From a constructive point of
view, such energy is not justified. 

There are several possible alternatives. One of them is not to consider
self-action (second alternative). A third alternative is to consider
only the energy contained in transversal (rotational) non-stationary
fields ($E_{\perp},B)$ (where $E_{\perp}=E-\left(-\nabla V_{C}\right)$,
being $V_{C}$ the Coulomb potential). In the first and third cases,
there will be a flow of energy associated to the EM waves each time
the entity is not in a stationary state. In the second case, energy
will correspond to interactions and the possibility of a quantum entity
to be isolated as the limit of constructively factual situations will
be doubted. One way or the other each view can be incorporated to
the formalism, which in turn will provide consistency requirements
that can be contrasted against experimental situations. Which alternative(s)
must be discarded is to be decided experimentally. The list of alternatives
will remain open until the occasion. We will see in Part II of this
work that the alternatives that present ``no self-actions in stationary
situations'' (second and third alternative) account for the energy
levels of the hydrogen atom. Yet, if we want to account for the photoelectric
effect or the Compton effect we need to have room for incoming or
outgoing energy and momentum associated to the EM fields, thus the
third alternative looks more promising.

It is worth noticing that in the statistical interpretation of quantum
mechanics, the energy of the field is not considered (no reasons offered)
avoiding in such way dealing with an infinite energy that corresponds
to the electrostatic (self) energy of a point charge. However, self
energies are not ruled out completely since they are needed (in the
standard argumentation) to account for the so called spin-orbit interaction.
As usual in the instrumentalist attitude, consistence is not a matter
of concern.

\part{Implementation: The hydrogen atom}

\section{Constructive ideas for Quantum Mechanics from Electromagnetism}

We borrow some basic ingredients from Maxwell's electrodynamics and
from traditional Quantum Mechanics. These elements will be combined
in dynamical equations in the next Section.
\begin{enumerate}
\item The atom is the microscopic entity susceptible of measurements. Proton
and electron within the atom do not have a sharp separable identity,
they are inferences (or ``shadows''). The identity of the atom consists
of charge, current, magnetisation(s) and masses.
\item The experiments of Uhlenbeck \& Goudsmidt and Zeeman as well as that
of Stern and Gerlach suggest the existence of intrinsic magnetisations.
We introduce it following the standard form relying in Pauli's seminal
work \citep{pauli1927quantum}. 
\item The interactions within the atom and of the atom with detectors are
electromagnetic.
\item Elementary entities such as the proton and the electron have no self-interactions.
There is no evidence regarding the existence of self-energies (constitutive
energies) at this level of description, correspondingly, they will
not be included.
\item The experimental detection of properties belonging to the shadows
within the atom depends on the actual possibility of coupling them
with measurement devices (usually based upon electromagnetic properties).
We will stay on the safe side and will not assume as measurable things
like the distance from proton to electron or the probability distribution
of the relative position between proton and electron. Such ``measurements''
remain in the domain of fantasy unless the actual procedure is offered
for examination. In this respect we severely depart from textbook
expositions of Quantum Mechanics.
\item The atom as an EM-entity is described with wave-functions. Any quantity
associated to an atom, such as its EM fields, are represented by operators
on the wave-function's space.
\end{enumerate}

\subsection{Expression of the electromagnetic potentials as integrated values
on operators }

\subsubsection{Wave-functions}

Two bodies that are spatially separated are idealised as represented
by a \emph{wave-function} which is the product of independent wave-functions
on different coordinates. For the case of a proton and an electron
we write $\Psi(x_{e},x_{p})=\psi_{e}(x_{e})\psi_{p}(x_{p})$ and call
it the \emph{electromagnetic limit}. In contrast, a single body with
an internal structure is represented by the product of a wave-function
associated to the body and another one representing the internal degrees
of freedom. For the hydrogen atom $\Phi(x_{r},x_{cm})=\phi_{r}(x_{r})\phi_{cm}(x_{cm})$
represents the \emph{atomic limit}. 

The election of the centre of mass, $cm$, to represent the position
of the body will be shown later to be consistent with the form of
the relational kinetic energy introduced. The variable $x_{r}$ ($r$
for \emph{relative} or \emph{relationa}l) stands for $x_{e}-x_{p}$.
Our constructive postulate is that, when proton and electron are conceived
as (spatially) separated entities, the laws of motion correspond to
electromagnetism and when conceived as a hydrogen atom, the usual
QM is recovered as a limit case\footnote{Conventional QM also assumes whenever necessary that $\psi_{CM}(x_{CM})$
corresponds to a point-like particle at rest.}. In fact, the present approach is broader than the usual QM in at
least two aspects: (a) The decomposition of the wave-function as a
product is not imposed, but rather regarded as a limit and (b) the
proton is regarded as something more than just the (point-charge)
source of electrostatic potential. Hence, starting from the laws of
EM, the structure of a general quantum theory for $|\Psi\rangle$
can be postulated as surpassing EM. Specialising this theory to the
atomic case, the QM of the hydrogen atom is obtained.

\subsubsection{Integrals on operators}

We will call the value of the integral 
\begin{equation}
\int d^{3}x_{a}\ d^{3}x_{b}\ \left(\Psi^{*}(x_{a},x_{b})(\hat{O}\Psi(x_{a},x_{b}))\right)\equiv\left\langle \Psi^{*}(\hat{O}\Psi)\right\rangle \equiv\left\langle \Psi^{*}|\hat{O}|\Psi)\right\rangle \label{eq:integrated-values}
\end{equation}
the \emph{integrated value} of the operator $\hat{O}$, where $\hat{O}$
is an operator that maps the wave-function $\Psi$ as $\hat{O}\Psi(x_{a},x_{b})=\Psi^{\prime}(x_{a},x_{b})$\footnote{We will drop the time-argument of the wave-function whenever possible
to lighten the notation.}.

\subsubsection{Physical Background}

In this section our goal is to develop an electromagnetic theory of
microscopic systems encompassing both QM and Maxwell's electrodynamics,
starting with the principles stated at the beginning of this Section
and in Part  of this work \citep{sola2024}. Hence, we will compute
Maxwell's EM interaction energy, either between the atom and external
EM-fields or between different fields arising within the atom. We
display first some classical results that are relevant for the coming
computations.

Maxwell's electromagnetic \emph{interaction energy} between entities
$1$ and $2$ reads:

\textbf{
\begin{equation}
{\cal E}=\int d^{3}x\left[\epsilon_{0}E_{1}\cdot E_{2}+\frac{1}{\mu_{0}}B_{1}\cdot B_{2}\right].\label{eq:EM-energy-1}
\end{equation}
}As such, this result is independent of the choice of gauge for the
electromagnetic potentials $A,V$. In the sequel, we will adopt the
Coulomb gauge whenever required.

The vector and scalar potentials associated to a charge-current density
$(Cq\rho,j)$ in a (electro- and magneto-) static situation are
\begin{eqnarray*}
\left(\frac{V}{C},A\right)(x,t) & = & \frac{\mu_{0}}{4\pi}\int d^{3}y\,\frac{(Cq\rho,j)(y,t)}{|x-y|}.
\end{eqnarray*}

The vector potential associated to a magnetisation $M$ at the coordinate
$y$ reads:
\begin{eqnarray*}
A_{M}(x,t) & = & \frac{\mu_{0}}{4\pi}\nabla_{x}\times\int d^{3}y\,\frac{M(y,t)}{|x-y|}.
\end{eqnarray*}
For potentials as defined in the Lorenz gauge there is a gauge transformation
that brings them to Coulomb form i.e., $\rho$ electrostatic and $\nabla\cdot A=0$.
Finally, electric and magnetic fields are obtained through Maxwell's
equations ${\displaystyle E=-\nabla V-\frac{\partial A}{\partial t}}$
and $B=\nabla\times A$.

\subsubsection{Electromagnetic static potentials}

We define the density within the atom as $\rho(x_{a},x_{b})=|\Psi(x_{a},x_{b})|^{2}$.
When appropriate, protonic and electronic densities may be defined
as\linebreak{}
 $\int d^{3}x_{b}\left|\Psi(x_{a},x_{b})\right|^{2}\equiv\rho_{a}(x_{a})$
for each component, $a=e,p$. In the electromagnetic limit of full
separation between electron and proton ($|\Psi\rangle=|\psi_{a}\rangle|\psi_{b}\rangle$),
with a wave-function for particle $b$ normalised to unity, we recover
$\rho_{a}(x_{a})=|\psi_{a}(x_{a})|^{2}$. Protonic and electronic
current-densities within the atom are defined similarly for $a=e,p$
by 
\[
j_{a}(x_{p},x_{e})=-\frac{i\hbar q_{a}}{2m_{a}}\left(\Psi^{*}(x_{p},x_{e})\left(\nabla_{a}\Psi(x_{p},x_{e})\right)-\Psi(x_{p},x_{e})\left(\nabla_{a}\Psi^{*}(x_{p},x_{e})\right)\right)
\]
with $q_{e}=-q_{p}$. 

The electrostatic potential reads 
\begin{eqnarray*}
V_{a}(x,t) & = & \frac{q_{a}}{4\pi\epsilon_{0}}\int\!d^{3}x_{a}\,d^{3}x_{b}\ \frac{1}{|x-x_{a}|}\left|\Psi(x_{a},x_{b})\right|^{2}=\frac{q_{a}}{4\pi\epsilon_{0}}\langle\Psi|\frac{1}{|x-x_{a}|}|\Psi\rangle.
\end{eqnarray*}
Moreover, 
\[
V_{a}(x,t)=\frac{q_{a}}{4\pi\epsilon_{0}}\int\!d^{3}x_{a}\ \frac{1}{|x-x_{a}|}\left(\int d^{3}x_{b}\left|\Psi(x_{a},x_{b})\right|^{2}\right)=\frac{q_{a}}{4\pi\epsilon_{0}}\int\!d^{3}x_{a}\ \frac{\rho_{a}(x_{a})}{|x-x_{a}|}
\]
showing that the electromagnetic structure is preserved in general,
not just in the limit of full separation.

Similarly, the vector potential (Coulomb gauge) associated to an intrinsic
microscopic magnetisation $M_{a}(x_{a},x_{b})=M_{a}(x_{a})\rho(x_{a},x_{b})$
reads: 
\begin{eqnarray*}
A_{M_{a}}(x,t) & = & \frac{\mu_{0}}{4\pi}\nabla_{x}\times\int d^{3}x_{b}\ d^{3}x_{a}\,\frac{M_{a}(x_{a},x_{b})}{|x-x_{a}|}\\
 & = & \frac{\mu_{0}}{4\pi}\nabla_{x}\times\int d^{3}x_{a}\,\frac{M_{a}(x_{a})}{|x-x_{a}|}\rho(x_{a})\\
 & = & \frac{\mu_{0}}{4\pi}\nabla_{x}\times\langle\Psi|\frac{M_{a}}{|x-x_{a}|}|\Psi\rangle
\end{eqnarray*}
(the index in $\nabla$ indicates the derivation variable) where $\Psi$
is the wave-function. In the electromagnetic limit, the vector potential
reads:
\begin{eqnarray*}
A_{M_{a}}(x,t) & = & \frac{\mu_{0}}{4\pi}\nabla_{x}\times\int\!d^{3}x_{a}\,\frac{M_{a}}{|x-x_{a}|}\left|\psi_{a}(x_{a})\right|^{2}\\
 & = & \frac{\mu_{0}}{4\pi}\nabla_{x}\times\langle\psi_{a}|\frac{M_{a}}{|x-x_{a}|}|\psi_{a}\rangle,
\end{eqnarray*}
which is the classical EM potential for a density $\left|\psi_{a}(x_{a})\right|^{2}$. 

For a static current we have 
\begin{eqnarray*}
A_{a}(x,t) & = & \frac{1}{2}\frac{\mu_{0}q_{a}}{4\pi m_{a}}\int\!d^{3}x_{a}\,d^{3}x_{b}\ \left(\Psi^{*}(x_{a},x_{b})\frac{1}{|x-x_{a}|}(-i\hbar\nabla_{a})\Psi(x_{a},x_{b})+c.c.\right).
\end{eqnarray*}
 Here we may also define 
\[
{\displaystyle \widetilde{j}_{a}(x_{a})=\frac{q_{a}}{2m_{a}}\int d^{3}x_{b}\ \left(\Psi^{*}(x_{a},x_{b})(-i\hbar\nabla_{a})\Psi(x_{a},x_{b})\right)+c.c.}
\]
 and set
\begin{eqnarray*}
A_{a}(x,t) & = & \frac{\mu_{0}}{4\pi}\int\!d^{3}x_{a}\ \left(\frac{\widetilde{j}_{a}(x_{a})}{|x-x_{a}|}\right).
\end{eqnarray*}

\subsubsection{On atomic magnetic moments}

We associate an intrinsic magnetisation to proton and electron, following
Pauli's original work \citep{pauli1927quantum}. A general form for
the intrinsic magnetisation for a microscopic constituent $a$ reads
\[
M_{a}=\frac{q_{a}}{m_{a}}S_{a}
\]
where $S_{a}={\displaystyle \frac{\hbar}{2}\sigma^{a}}$ and $\sigma^{a}$
are the Pauli matrices for constituent $a$.

\subsection{Variational formulation}

Both Newtonian mechanics \citep{arno89} and Maxwell's electrodynamics
\citep{solari2022symmetries} can be reformulated as variational theories,
satisfying the \emph{least action principle. }The action (or action
integral) is the integral of the Lagrangian over time. From Part I,
{\footnotesize{}
\begin{equation}
{\cal L}=\int d^{3}y\,\left[\frac{1}{2}\left(\Psi^{*}\left(i\hbar\dot{\Psi}\right)+(i\hbar\dot{\Psi})^{*}\Psi\right)+\frac{\hbar^{2}}{2m}\Psi^{*}T\Psi+\frac{1}{2}\sum_{i\ne j}\left(\frac{1}{\mu_{0}}B_{i}\cdot B_{j}-\epsilon_{0}E_{i}\cdot E_{j}\right)\right]\label{eq:langrangianFinal}
\end{equation}
}with $i,j\in\left\{ e,p,L\right\} $ which stand for electron, proton
and Laboratory. The symbol $T$ stands for kinetic energy and its
form will be presented later. Actually ``$d^{3}y$'' is a symbol
indicating that the integral is performed over all non-temporal arguments
of the wave-function, which in the case of the hydrogen atom will
be at least six.

In this Part of the work we are interested in the internal energies
associated to stationary states of the microscopic entity (the Hydrogen
atom). In Part I we assumed that the microscopic entity did not have
an internal structure and therefore there was no ``self-energy''.
In the present situation the atomic energy will have an internal part
and an interaction part, as implied by eq.\eqref{eq:langrangianFinal}.

We propose therefore that the stationary states of the hydrogen atom
have an associated wave-function $\Psi_{\omega}(x_{e},x_{p})$, where
the indices $e$ and $p$ correspond to the inferred electron and
proton, whose unifying entity is the atom. $x$ represents appropriate
arguments of the wave-function, satisfying
\begin{equation}
\left(\hbar\omega\Psi_{\omega}\right)-\frac{\hbar^{2}}{2m}T\Psi_{\omega}-\hat{H}\Psi_{\omega}=0\label{eq:eigenvalue-1}
\end{equation}
(we drop the index $\omega$ in the sequel) where {\footnotesize{}
\[
\left\langle \delta\Psi\hat{H}_{EM}\Psi\right\rangle =\delta_{\Psi^{*}}\left\langle \frac{1}{\mu_{0}}\left(B_{L}\cdot\left(B_{e}+B_{p}\right)+B_{e}\cdot B_{p}\right)+\epsilon_{0}\left(E_{L}\cdot\left(E_{e}+E_{p}\right)+E_{e}\cdot E_{p}\right)\right\rangle 
\]
}the variation of the wave-function is $\delta\Psi$ and $\left\langle \cdots\right\rangle $
indicates integration over all coordinates as in eq.\eqref{eq:integrated-values}.

\section{Contributions to the integrated value of $H$}

\subsection{Relational kinetic energy\label{subsec:Relational-kinetic-energy}}

In terms of classical mechanics, the kinetic energy associated to
the hydrogen atom can be organised as follows. To fix ideas let index
$b$ correspond to the proton and $a$ to the electron. The quantum-mechanical
wave-function is $\Psi(x_{a},x_{b})$. The center-of-mass coordinate
(relative to some external reference such as ``the laboratory'')
and the relative coordinate (which is an invariant of the description,
independent of an external reference) can be defined as usual, satisfying
\begin{eqnarray*}
x_{cm} & = & \frac{m_{a}x_{a}+m_{b}x_{b}}{m_{a}+m_{b}}\\
x_{r} & = & x_{a}-x_{b}\\
x_{a} & = & x_{cm}+x_{r}\frac{m_{b}}{m_{a}+m_{b}}\\
x_{b} & = & x_{cm}-x_{r}\frac{m_{a}}{m_{a}+m_{b}}.
\end{eqnarray*}
The wave-function and the gradient (we also make use of $p_{x}=-i\hbar\nabla_{x})$
read:
\begin{eqnarray*}
\Psi(x_{a},x_{b}) & = & \Psi(x_{cm}+x_{r}\frac{m_{b}}{m_{a}+m_{b}},x_{cm}-x_{r}\frac{m_{a}}{m_{a}+m_{b}})=\Phi(x_{cm},x_{r})\\
\nabla_{cm} & = & \nabla_{a}+\nabla_{b}\\
\nabla_{r} & = & \frac{m_{b}}{m_{a}+m_{b}}\nabla_{a}-\frac{m_{a}}{m_{a}+m_{b}}\nabla_{b}\\
\Phi(x_{cm},x_{r}) & = & \phi(\frac{m_{a}x_{a}+m_{b}x_{b}}{m_{a}+m_{b}},x_{a}-x_{b})=\Psi(x_{a},x_{b})\\
\nabla_{a} & = & \nabla_{cm}\frac{m_{a}}{m_{a}+m_{b}}+\nabla_{r}\\
\nabla_{b} & = & \nabla_{cm}\frac{m_{b}}{m_{a}+m_{b}}-\nabla_{r}
\end{eqnarray*}
Hence, 
\begin{eqnarray*}
T_{a}+T_{b}=\frac{p_{a}^{2}}{2m_{a}}+\frac{p_{b}^{2}}{2m_{b}} & = & -\hbar^{2}\left(\nabla_{cm}^{2}\frac{1}{2(m_{a}+m_{b})}+\nabla_{r}^{2}\frac{(m_{a}+m_{b})}{2m_{a}m_{b}}\right)\\
 & = & \frac{-\hbar^{2}}{2(m_{a}+m_{b})}\nabla_{cm}^{2}+\frac{-\hbar^{2}}{2m_{r}}\nabla_{r}^{2}=T_{cm}+T_{r}.
\end{eqnarray*}
The associated integrated values read: 
\begin{eqnarray*}
\mathcal{E}_{k}=\langle\Psi|T_{a}|\Psi\text{\ensuremath{\rangle}}+\langle\Psi|T_{b}|\Psi\text{\ensuremath{\rangle}} & = & \langle\Phi|T_{cm}|\Phi\ensuremath{\rangle}+\langle\Phi|T_{r}|\Phi\text{\ensuremath{\rangle}}
\end{eqnarray*}

\subsection{Relative electrostatic potential energy }

The Coulomb electrostatic interaction energy for an atomic density
$\left|\Psi(x_{a},x_{b})\right|^{2}$ reads:
\begin{eqnarray*}
{\cal E}_{C}^{x_{a},x_{b}} & = & \epsilon_{0}\int d^{3}x\,E_{1}\cdot E_{2}\left|\Psi(x_{a},x_{b})\right|^{2}=-\epsilon_{0}\int d^{3}x\,E_{1}\cdot\nabla V_{2}\left|\Psi(x_{a},x_{b})\right|^{2}\\
 & = & \int d^{3}x\,\epsilon_{0}\left[-\nabla\cdot\left(E_{1}V_{2}\right)+\left(\nabla\cdot E_{1}\right)V_{2}\right]\left|\Psi(x_{a},x_{b})\right|^{2}.
\end{eqnarray*}
The full energy is given by averaging over the wave function $\Psi$.
The first contribution vanishes when integrated over all space by
Gauss theorem, and the second term can be recast as $-\epsilon_{0}(\nabla\cdot\nabla V_{1})V_{2},$
namely
\begin{eqnarray*}
{\cal E}_{C} & = & \int d^{3}x_{a}\ d^{3}x_{b}\ \left|\Psi(x_{a},x_{b})\right|^{2}\int d^{3}x\,\epsilon_{0}\left(\frac{q_{a}q_{b}}{\left(4\pi\epsilon_{0}\right)^{2}}\right)\left(-\frac{1}{|x-x_{b}|}\Delta\frac{1}{|x-x_{a}|}\right)\\
 & = & {\displaystyle -\frac{e^{2}}{4\pi\epsilon_{0}}}\int d^{3}x_{e}\ d^{3}x_{p}\ \left|\Psi(x_{e},x_{p})\right|^{2}\frac{1}{|x_{e}-x_{p}|}=-\langle\Psi|\frac{e^{2}}{4\pi\epsilon_{0}}\frac{1}{|x_{e}-x_{p}|}|\Psi\rangle
\end{eqnarray*}
where $q_{a}=e=-q_{b}$ and $e$ is the electron charge (negative),
i.e., $a$ is the electron, $b$ the proton. In the electromagnetic
limit it corresponds to the potential energy between two separate
charge densities $\rho_{k=}|\psi_{k}(x_{k})|^{2}$ of opposite sign,
corresponding to Maxwell's equation $q_{k}\rho_{k}=\epsilon_{0}\nabla\cdot E_{k}$. 

\subsection{Interaction of relative current with external magnetic field}

For this interaction \emph{relative current} means the currents ($j_{a},j_{b}$)
of both electron and proton regarding their motion relative to the
(source of) external field. Recall that we associate ${\displaystyle j=qv=\frac{q}{m}p=\frac{q}{m}(-i\hbar\nabla)}$,
i.e., current as a property of charge in (perceived) motion. The electromagnetic
energy in the external field $B(x)$ is {\footnotesize{}
\begin{eqnarray*}
\mathcal{E}_{jB} & \!\!= & \!\!\frac{1}{\mu_{0}}\int d^{3}x_{a}\ d^{3}x_{b}\ \Psi(x_{a},x_{b})^{*}\!\!\int d^{3}x\,B(x)\cdot\nabla\times\left(A_{a}+A_{b}\right)\Psi(x_{a},x_{b})\\
 & \!\!= & \!\!\frac{1}{2}\frac{1}{\mu_{0}}\frac{\mu_{0}}{4\pi}\int d^{3}x_{a}\ d^{3}x_{b}\ \Psi(x_{a},x_{b})^{*}\!\!\int d^{3}x\,B(x)\cdot\nabla\times\left(\frac{1}{|x-x_{a}|}\frac{q_{a}}{m_{a}}p_{a}\right)\Psi(x_{a},x_{b})+c.c.\\
 & \!\!+ & \!\!\frac{1}{2}\frac{1}{\mu_{0}}\frac{\mu_{0}}{4\pi}\int d^{3}x_{a}\ d^{3}x_{b}\ \Psi(x_{a},x_{b})^{*}\!\!\int d^{3}x\,B(x)\cdot\nabla\times\left(\frac{1}{|x-x_{b}|}\frac{q_{b}}{m_{b}}p_{b}\right)\Psi(x_{a},x_{b})+c.c.
\end{eqnarray*}
}Having in mind the (normal) Zeeman effect, we perform an explicit
calculation using that the vector potential for the approximately
constant external field $B$ reads $B\equiv\nabla\times A_{L}=\nabla\times\left(\chi(x)\frac{1}{2}x\times B\right)$,
with $\chi$ a smoothed version of the characteristic function for
the experiment's region. We place the origin of coordinates inside
the apparatus and assume the characteristic function to be 1 inside
a macroscopically large ball $K$ around the origin, so that 
\[
\int_{K\times K}dx_{a}\ dx_{b}\ |\Psi(x_{a},x_{b})|^{2}\simeq1
\]
(i.e., the atom is inside the measurement device). In such a case,
{\footnotesize{}
\begin{eqnarray*}
{\cal E}_{jB} & \sim & \frac{1}{2}\int d^{3}x_{a}\ d^{3}x_{b}\ \Psi(x_{a},x_{b})^{*}\!\!\left(\frac{1}{2}x_{a}\times B\cdot\frac{q_{a}}{m_{a}}p_{a}+\frac{1}{2}x_{b}\times B\cdot\frac{q_{b}}{m_{b}}p_{b}\right)\Psi(x_{a},x_{b})+c.c.\\
 & = & -\frac{1}{2}\int d^{3}x_{a}\ d^{3}x_{b}\ \Psi(x_{a},x_{b})^{*}\,\frac{1}{2}B\cdot\left(x_{a}\times\frac{q_{a}}{m_{a}}p_{a}+x_{b}\times\frac{q_{b}}{m_{b}}p_{b}\right)\Psi(x_{a},x_{b})+c.c.\\
 & = & -\frac{e}{2m_{r}}\langle B\cdot\left(\frac{m_{r}}{m_{e}+m_{p}}x_{r}\times p_{CM}+x_{CM}\times p_{r}+\frac{m_{p}-m_{e}}{m_{p}+m_{e}}L_{r}\right)\rangle\\
 & \sim & -\left(\frac{e}{2m_{e}}+O(\frac{m_{e}}{m_{p}})\right)\langle\Psi|B\cdot L_{e}|\Psi\rangle
\end{eqnarray*}
}where in the last line we have switched to $CM,r$ coordinates and
used that $\langle p_{r}\rangle=\langle p_{CM}\rangle=0$ and $[x_{r},\nabla_{CM}]=[x_{CM},\nabla_{r}]=0$.

Further, the argument of the above integral reads,
\[
\nabla\times A_{L}\cdot\nabla\times A_{a}=\nabla\cdot\left(A_{L}\times\left(\nabla\times A_{a}\right)\right)+A_{L}\cdot\nabla\times\left(\nabla\times A_{a}\right).
\]
The first term vanishes by Gauss theorem. The experimental conditions
assume that the atom lies well inside the region of magnetic field
and that the border effects ${\displaystyle \nabla\chi(x-x_{CM})\times\left(\frac{1}{2}x\times B\right)}$
can be disregarded. Moreover, $\nabla\times\left(\nabla\times A_{a}\right)=\nabla\left(\nabla\cdot A_{a}\right)-\Delta A_{a}$.
For the first term, we integrate over $x_{a},x_{b}$, using that\newline
${\displaystyle \nabla_{x}\frac{1}{|x-x_{a}|}=-\nabla_{x_{a}}\frac{1}{|x-x_{a}|}}$,
arriving to the integral of \newline ${\displaystyle \frac{1}{|x-x_{a}|}\nabla_{x_{a}}\!\cdot\!\left(\Psi(x_{a},x_{b})^{*}\frac{q_{a}}{m_{a}}p_{a}\Psi(x_{a},x_{b})\right)+c.c.}$,
which is zero for an atom in a stationary state, since by the continuity
equation it corresponds to the time-derivative of the atomic charge
density. For the other term, recall that \newline ${\displaystyle -\Delta\frac{1}{|x-x_{a}|}=4\pi\delta(x-x_{a})}$.
Hence, {\footnotesize{}
\begin{eqnarray*}
{\cal E}_{jB} & \!\!= & \!\!\frac{1}{2}\int d^{3}x_{a}\ d^{3}x_{b}\ \Psi(x_{a},x_{b})^{*}\!\!\int d^{3}x\,A_{L}\cdot\left(\delta(x-x_{a})\frac{q_{a}}{m_{a}}p_{a}+\delta(x-x_{b})\frac{q_{b}}{m_{b}}p_{b}\right)\Psi(x_{a},x_{b})\\
 &  & +c.c.\\
 & \!\!= & \!\!\frac{1}{2}\int d^{3}x_{a}\ d^{3}x_{b}\ \Psi(x_{a},x_{b})^{*}\!\!\left(A_{L}(x_{a})\cdot\frac{q_{a}}{m_{a}}p_{a}+A_{L}(x_{b})\cdot\frac{q_{b}}{m_{b}}p_{b}\right)\Psi(x_{a},x_{b})+c.c.\\
 & \!\!\sim & \!\!\frac{1}{2}\int d^{3}x_{a}\ d^{3}x_{b}\ \Psi(x_{a},x_{b})^{*}\!\!\left(\frac{1}{2}x_{a}\times B\cdot\frac{q_{a}}{m_{a}}p_{a}+\frac{1}{2}x_{b}\times B\cdot\frac{q_{b}}{m_{b}}p_{b}\right)\Psi(x_{a},x_{b})+c.c.
\end{eqnarray*}
}assuming that the atomic wave-function is negligibly small outside
the border of the experiment's region. We now move to center-of-mass
and relative coordinates, assuming further that the atom as a whole
is at rest (at the centre of mass location) during the experiment
(i.e., $\langle\Psi|p_{CM}|\Psi\rangle=0$, $p_{r}\sim p_{e}$, $x_{r}\sim x_{e}$
and $\frac{m_{p}-m_{e}}{m_{p}+m_{e}}\sim1$). Hence,{\small{}
\begin{eqnarray*}
{\cal E}_{jB} & \sim & \frac{1}{2}\,\left\{ \left\langle \Psi(x_{a},x_{b})^{*}\!\left(\frac{1}{2}(x_{a}-x_{CM})\times B\cdot\frac{q_{a}}{m_{a}}\right)\!\Psi(x_{a},x_{b})\right\rangle \right.\\
 &  & +\left.\left\langle \Psi(x_{a},x_{b})^{*}\!\left(\frac{1}{2}(x_{b}-x_{CM})\times B\cdot\frac{q_{b}}{m_{b}}p_{b}\right)\!\Psi(x_{a},x_{b})\right\rangle \right\} +c.c.\\
 & = & \frac{1}{2}\left(-\frac{e}{2(m_{p}+m_{e})}\right)\left\langle \Psi(x_{CM},x_{r})^{*}\,B\cdot\left(\frac{m_{p}}{m_{e}}-\frac{m_{e}}{m_{p}}\right)L_{r}\Psi(x_{CM},x_{r})\right\rangle +c.c.\\
 & = & -\frac{e}{2m_{r}}\frac{m_{p}-m_{e}}{m_{p}+m_{e}}\langle\Psi|B\cdot L_{r}|\Psi\rangle\sim-\left(\frac{e}{2m_{e}}+O(\frac{m_{e}}{m_{p}})\right)\langle\Psi|B\cdot L_{e}|\Psi\rangle
\end{eqnarray*}
}{\small\par}

\subsection{Interaction of spin(s) with external magnetic field\label{subsec:SB}}

We regard spin as an intrinsic magnetisation ${\displaystyle \frac{q_{a}}{m_{a}}S_{a}}$,
leading to a vector potential 
\begin{eqnarray}
A_{s}(x,t) & = & \frac{\mu_{0}}{4\pi}\nabla\times\langle\Psi|\left(\frac{q_{a}}{m_{a}}\frac{S_{a}}{|x-x_{a}|}-\frac{q_{a}}{m_{b}}\frac{S_{b}}{|x-x_{b}|}\right)|\Psi\rangle\label{eq:atomicA}
\end{eqnarray}
whose interaction energy with an external magnetic field $B(x,t)$
reads
\begin{eqnarray*}
\mathcal{E}_{SB} & = & \frac{1}{\mu_{0}}\int d^{3}x\ B(x,t)\cdot\nabla\times A_{s}(x,t)\\
 & = & \frac{e}{4\pi}\langle\Psi|\int d^{3}x\ B(x,t)\cdot\left(\nabla\times\nabla\times\left(\frac{1}{m_{e}}\frac{S_{e}}{|x-x_{e}|}-\frac{1}{m_{p}}\frac{S_{p}}{|x-x_{p}|}\right)\right)|\Psi\rangle\\
 & = & \frac{e}{4\pi}\langle\Psi|\int d^{3}x\ B(x,t)\cdot\left[\nabla\left(\nabla\cdot\left(\frac{1}{m_{e}}\frac{S_{e}}{|x-x_{e}|}-\frac{1}{m_{p}}\frac{S_{p}}{|x-x_{p}|}\right)\right)\right]|\Psi\rangle\\
 & - & \frac{e}{4\pi}\langle\Psi|\int d^{3}x\ B(x,t)\cdot\left(\Delta\left(\frac{1}{m_{e}}\frac{S_{e}}{|x-x_{e}|}-\frac{1}{m_{p}}\frac{S_{p}}{|x-x_{p}|}\right)\right)|\Psi\rangle
\end{eqnarray*}
Since $\nabla\cdot B=0$, the first contribution integrates to zero
by Gauss theorem since $B\cdot\nabla\phi=\nabla\cdot(B\phi)-\phi\nabla\cdot B$.
Hence,
\begin{eqnarray*}
\mathcal{E}_{SB} & = & e\langle\Psi|\left(\frac{1}{m_{e}}B(x_{e},t)\cdot S_{e}-\frac{1}{m_{p}}B(x_{p},t)\cdot S_{p}\right)|\Psi\rangle
\end{eqnarray*}
thus completing the description of Zeeman effect. In the atomic limit
the usual expression is recovered. 

\paragraph{Stern-Gerlach effect.}

All forces, including the Lorentz force \citep{solari2022symmetries},
are obtained from actions as the response of the action integral to
a variation of the relative position of the interacting bodies.  Displacing
by $\delta x$ the position of the field $B$ one gets 
\begin{eqnarray*}
\delta\mathcal{E}_{SB} & = & e\langle\Psi|\left(\frac{1}{m_{e}}B(x_{e}+\delta x,t)\cdot S_{e}-\frac{1}{m_{p}}B(x_{p}+\delta x,t)\cdot S_{p}\right)|\Psi\rangle=\delta x\cdot F
\end{eqnarray*}
 and then $F\sim\langle\Psi|\nabla\left(\frac{e}{m_{e}}B(x,t)\cdot S\right)|\Psi\rangle$.
The force is nonzero only for spatially varying magnetic fields as
observed in the experiment \citet{bauer2023sterngerlach}.

\subsection{Spin-orbit interaction\label{subsec:SO}}

If we follow the standard discourse of Quantum physics, spin-orbit
represents mainly the interaction between the electron's orbit and
spin. Such heuristic approach would break one of our fundamental propositions:
all the energies in the hydrogen atom are interaction energies between
proton and electron, no self energy is involved. A relational view
is actually forced to recognise that the orbit of the electron is
a motion relative to the proton. Hence, only their relative velocity
can matter. Such observation does not solve our problem, ... but let
us follow its lead.

The coupling of spin and relative current has two parts, namely the
interaction of proton spin ${\displaystyle \frac{-e}{m_{p}}}S_{p}$
with the relative current ${\displaystyle \frac{ep_{r}}{m_{r}}}$
as perceived by the proton and the corresponding interaction of electron
spin ${\displaystyle \frac{e}{m_{e}}}S_{e}$ with relative current
${\displaystyle \frac{-e(-p_{r})}{m_{r}}}$ as perceived by the electron.
${\displaystyle m_{r}=\frac{m_{e}m_{p}}{m_{e}+m_{p}}}$ stands for
the reduced mass and ${\displaystyle e\left(\frac{p_{e}}{m_{e}}-\frac{p_{p}}{m_{p}}\right)=e\frac{p_{r}}{m_{r}}}$.
The magnetic field operator associated to the relative current as
seen by the proton is hence the curl of the vector potential operator
associated to that current (recall that the energy contribution is
the integrated value of the operators over the wave-function), and
correspondingly for the current perceived by the electron: 
\begin{eqnarray}
\hat{B}_{jp}(x)|\Psi\rangle & = & \frac{\mu_{0}e}{4\pi}\nabla_{x}\times\left({\displaystyle \frac{1}{|x-x_{p}|}\left(\frac{p_{e}}{m_{e}}-\frac{p_{p}}{m_{p}}\right)}\right)|\Psi\rangle\nonumber \\
 & = & \frac{\mu_{0}e}{4\pi}\nabla_{x}{\displaystyle \frac{1}{|x-x_{p}|}\times\left(\frac{p_{e}}{m_{e}}-\frac{p_{p}}{m_{p}}\right)}|\Psi\rangle\label{eq:SO-fields}\\
\hat{B}_{je}(x)|\Psi\rangle & = & \frac{\mu_{0}e}{4\pi}{\displaystyle \nabla_{x}{\displaystyle \frac{1}{|x-x_{e}|}\times\left(\frac{p_{e}}{m_{e}}-\frac{p_{p}}{m_{p}}\right)}}|\Psi\rangle.\nonumber 
\end{eqnarray}
 The magnetic field operators associated to the spin are:
\begin{eqnarray*}
\hat{B}_{se}(x) & = & \frac{\mu_{0}e}{4\pi}\nabla_{x}\times\left(\nabla_{x}\times\left(\frac{1}{|x-x_{e}|}\frac{S_{e}}{m_{e}}\right)\right)\\
\hat{B}_{sp}(x) & = & -\frac{\mu_{0}e}{4\pi}\nabla_{x}\times\left(\nabla_{x}\times\left(\frac{1}{|x-x_{p}|}\frac{S_{p}}{m_{p}}\right)\right)
\end{eqnarray*}
and the energy contribution is 
\begin{eqnarray*}
\mathcal{E}_{SO} & = & \kappa\frac{1}{\mu_{0}}\langle\Psi|\int d^{3}x\ \hat{B}_{se}(x)\cdot\hat{B}_{jp}(x)+\hat{B}_{sp}(x)\cdot\hat{B}_{je}(x)|\Psi\rangle
\end{eqnarray*}
where $\kappa$ is a numerical constant that needs to be determined.
We transform the spin field operators as 

{\small{}
\begin{eqnarray}
\hat{B}_{se}(x)\cdot\hat{B}_{jp}(x) & \!\!= & \!\!\frac{\mu_{0}e}{4\pi m_{e}}\hat{B}_{jp}(x)\cdot\left(\nabla\left(\nabla\cdot\frac{S_{e}}{|x-x_{e}|}\right)-\Delta\left(\frac{S_{e}}{|x-x_{e}|}\right)\right)\nonumber \\
 & \!\!= & \!\!\frac{\mu_{0}e}{4\pi m_{e}}\!\!\left(\nabla\cdot\left(\hat{B}_{jp}(x)\left(\nabla\cdot\frac{S_{e}}{|x-x_{e}|}\right)\right)-\hat{B}_{jp}(x)\cdot\Delta\left(\frac{S_{e}}{|x-x_{e}|}\right)\right)\nonumber \\
 & \!\!= & \!\!\frac{\mu_{0}e}{4\pi m_{e}}\hat{B}_{jp}(x)\cdot\left(4\pi S_{e}\delta(x-x_{e})\right)\nonumber \\
\hat{B}_{sp}(x)\cdot\hat{B}_{je}(x) & \!\!= & \!\!-\frac{\mu_{0}e}{4\pi m_{p}}\hat{B}_{je}(x)\cdot\left(4\pi S_{p}\delta(x-x_{p})\right)\label{eq:SO-segunda}
\end{eqnarray}
}using Gauss theorem and that $\nabla\cdot B=0$. Performing the $x-$integral
first, we obtain 
\begin{eqnarray}
\mathcal{E}_{SO} & = & -\kappa\frac{\mu_{0}e^{2}}{4\pi}\langle\Psi|\left(\frac{S_{e}}{m_{e}}+\frac{S_{p}}{m_{p}}\right)\cdot\left({\displaystyle \frac{-\left(x_{p}-x_{e}\right)}{|x_{e}-x_{p}|^{3}}\times\frac{p_{r}}{m_{r}}}\right)|\Psi\rangle\nonumber \\
 & = & -\kappa\frac{\mu_{0}e^{2}}{4\pi}\langle\Psi|\left(\frac{S_{e}}{m_{e}}+\frac{S_{p}}{m_{p}}\right)\cdot\left({\displaystyle \frac{L_{r}}{m_{r}|x_{e}-x_{p}|^{3}}}\right)|\Psi\rangle,\label{eq:SO-final}
\end{eqnarray}
where $L_{r}=x_{r}\times p_{r}$ is the relative angular momentum
operator. 

Let us complete the expression determining the value of $\kappa$
before we turn back to the question that opens this subsection. The
expression for the electromagnetic energy (Equation \ref{eq:EM-energy-1})
was introduced by Maxwell considering the kind of interactions known
at his time. It is symmetric in the indexes of the two interacting
systems, hence, if ${\cal P}(12)$ is a permutation of indexes, the
energy can be written as:

{\small{}
\[
{\cal E}_{SO}=\frac{1}{2}\left\{ \int d^{3}x\left[\epsilon_{0}E_{1}\cdot E_{2}+\frac{1}{\mu_{0}}B_{1}\cdot B_{2}\right]+{\cal P}(12)\int d^{3}x\left[\epsilon_{0}E_{1}\cdot E_{2}+\frac{1}{\mu_{0}}B_{1}\cdot B_{2}\right]\right\} .
\]
}We may call the first integral the way in which system one acts upon
system two and the second integral is the reciprocal action. The energy
can then be obtained by first establishing one action and next ``symmetrising''.
The action of ${\cal P}(12)$ is merely changing the point of view
in a somewhat arbitrary form, while the imposition of symmetry removes
the arbitrariness since the acting group is a group of arbitrariness
(see \citep{sola18b}). The operation of taking two different view
points corresponds to operating with ${\cal P}(ep)$ as it can be
easily verified in all the previous expressions. Hence, unless ${\displaystyle \kappa=\frac{1}{2}}$
we would not be counting the interaction properly.

It remains to show that we are dealing with an interaction between
two different entities and not an internal interaction. Consider our
final expression, Eq. \eqref{eq:SO-final} and write it as:
\begin{eqnarray*}
{\cal E_{SO}} & = & -\frac{1}{2}\frac{\mu_{0}e^{2}}{4\pi}\langle\Psi|\left(\frac{S_{e}}{m_{e}}+\frac{S_{p}}{m_{p}}\right)\cdot\left({\displaystyle \frac{(x_{e}-x_{p})\times(v_{e}-v_{p})}{|x_{e}-x_{p}|^{3}}}\right)|\Psi\rangle\\
 & = & \frac{1}{2}\int\!d^{3}x_{e}d^{3}x_{p}\left\{ \!\Psi^{\dagger}\left({\displaystyle \frac{(x_{e}-x_{p})}{|x_{e}-x_{p}|^{3}}}\right)\cdot\frac{e^{2}}{4\pi\epsilon_{0}C^{2}}\left(\frac{S_{e}\times v_{e}}{m_{e}}-\frac{S_{p}\times v_{p}}{m_{p}}\right)\Psi\!\right\} \\
 &  & -\frac{1}{2}\int d^{3}x_{e}d^{3}x_{p}\,\left\{ \Psi^{\dagger}\left({\displaystyle \frac{(x_{e}-x_{p})}{|x_{e}-x_{p}|^{3}}}\right)\cdot\frac{\mu_{0}e^{2}}{4\pi}\left(\frac{S_{e}\times v_{p}}{m_{e}}-\frac{S_{p}\times v_{e}}{m_{p}}\right)\Psi\right\} 
\end{eqnarray*}
($\Psi^{\dagger}$ stands for the row matrix that is the transpose
and complex conjugate of $\Psi$). Consider further the case $\Psi=\psi_{e}\psi_{p}$,
i.e., the case when $\Psi=\psi_{e}\psi_{p}$ and in addition the distance
$|x_{e}-x_{p}|$ is macroscopic. In such a case we can consider that
the variation of the relative position with respect to $\langle x_{e}-x_{p}\rangle$
is negligible in front of the macroscopic distance. The main contribution
of the first term reads 
\[
\sim\frac{1}{2}{\displaystyle \frac{1}{4\pi\epsilon_{0}C^{2}}\frac{(x_{p}-x_{e})}{|x_{e}-x_{p}|^{3}}\cdot\,\int d^{3}x_{e}\,\left(e\psi_{e}^{\dagger}\frac{S_{e}\times v_{e}}{m_{e}}\psi_{e}\right)\int d^{3}x_{p}}\left(-e|\psi_{p}|^{2}\right),
\]
which represents the interaction between an electric dipole\\
 $\int d^{3}x_{e}\,\left(e\psi_{e}^{\dagger}\frac{S_{e}\times v_{e}}{m_{e}}\psi_{e}\right)$,
located in the electron (now approximated as a point) and the proton
(approximated as a point in front of the macroscopic distance). This
view is compatible with the idea that magnetic dipoles in motion produce
electric dipole fields. Up to a certain point, $\,\left(e\psi_{e}^{\dagger}\frac{S_{e}\times v_{e}}{m_{e}}\psi_{e}\right)$
represents a density of electric dipoles and $-e|\psi_{p}|^{2}$ a
density of charge. The difference with such densities is that it is
not possible to limit the interaction to ``part of the electron''
or ``part of a proton'', hence a integration over full space is
always mandatory., thus the view of $\left(e\psi_{e}^{\dagger}\frac{S_{e}\times v_{e}}{m_{e}}\psi_{e}\right)$
as density of classical dipoles is only an analogy, it leaves aside
the unity of the electron.

Actually, the dominant term in spin-orbit interactions corresponds
to the electric dipole of the electron acting upon the proton. Thus,
what was described as an interaction between the electron and its
own orbit is now identified as the action of the electric dipole associated
with the moving electron with the proton charge. An equivalent contribution
appears with the action of ${\cal P}(ep)$. The last two terms are
magnetic field interactions between proton and electron.

The expression \ref{eq:SO-final} corresponds well with the final
expression in textbook derivations. However, we have not resorted
to analogy, nor have we patched this or any other expression with
gyromagnetic numbers and have not further patched the expression with
``relativistic corrections'' (such as Thomas' correction) in the
need to agree with experiments. The derivation of the spin-orbit contribution
highlights the differences between the utilitarian/instrumentalist
and the old style approach.

\paragraph{Related experiments }

One of the best known results of the atomic limit was the prediction
of the hydrogen spectroscopic lines with Schrödinger's wave equation,
providing a full theoretical expression of the Rydberg constant. The
transition $n=2$ to $n=1$ \citep{kramida23} is reported as the
spectroscopic line at $\lambda=1215.6699\mathring{A}$. The line-width
allows for a calculated resolution into $\lambda_{1}=1215.668237310\mathring{A}$
and $\lambda_{2}=1215.673644608\mathring{A}$ The state $n=2$ is
an octuplet, partially resolved by the spin-orbit Hamiltonian of Section
\ref{subsec:SO} into a quadruplet and a doublet for $l=1$ and two
unresolved states with $l=0.$ The difference between the two energy
levels is calculated as 
\begin{eqnarray*}
\Delta E_{SO} & = & \frac{3}{2}\hbar^{2}\frac{1}{24a_{0}^{3}}\left(\frac{1}{2}\frac{\mu_{0}e^{2}}{4\pi m_{e}^{2}}\right)\frac{m_{e}}{m_{r}}\\
 & = & \frac{3}{96}\frac{m_{e}}{m_{r}}\alpha^{4}m_{e}C^{2}\\
 & = & 7.259023470408092\cdot10^{-24}J
\end{eqnarray*}
while the calculated experimental energy difference amounts to\linebreak{}
 ${\displaystyle \Delta E_{exp}=hC(\frac{1}{\lambda_{1}}-\frac{1}{\lambda_{2}})=7.26816814178113}\cdot10^{-24}J.$
The contribution consists of: ${\displaystyle \frac{3}{2}\hbar^{2}}$
coming from the level difference in $\langle L\cdot S\rangle$ between
$j=3/2$ and $j=1/2$ (for $l=1$ and $s=1/2$), ${\displaystyle \frac{1}{24a_{0}^{3}}=\langle R_{21}(r)|\frac{1}{r^{3}}|R_{21}(r)\rangle}$
for the associated Hydrogen $(n,l)$-radial wave-function, $a_{0}={\displaystyle \frac{\hbar}{\alpha m_{e}C}}$
is the Bohr radius, $m_{e}$ is the electron mass and ${\displaystyle \alpha=\frac{\mu_{0}e^{2}C}{4\pi\hbar}}$
is the fine-structure constant. We have disregarded the nuclear spin
contribution. 

\subsection{Spin-spin interaction (part of hyperfine structure)}

The interaction between spins within the atom uses an expression for
the vector potential corresponding to that in Section \ref{subsec:SB}
and reads
\begin{eqnarray*}
\mathcal{E}_{SS} & = & \frac{1}{\mu_{0}}\langle\Psi|\int d^{3}x\ B_{e}(x,t)\cdot B_{p}(x,t)|\Psi\rangle\\
B_{e} & = & \nabla\times\left(\nabla\times\left(\frac{e}{m_{e}}\frac{\mu_{0}}{4\pi}\,\frac{S_{e}(x_{e})}{|x-x_{e}|}\right)\right)\\
B_{p} & = & \nabla\times\left(\nabla\times\left(\frac{-e}{m_{p}}\frac{\mu_{0}}{4\pi}\,\frac{S_{p}(x_{p})}{|x-x_{p}|}\right)\right)
\end{eqnarray*}
To proceed with the calculation we need to break the $e-p$ symmetry,
namely $\mathcal{E}_{SS}=\frac{1}{2}\left(\mathcal{E}_{SS}^{e\to p}+\mathcal{E}_{SS}^{p\to e}\right)$.
We pick the electron to proceed with the calculation, applying Gauss
theorem again as in Section \ref{subsec:SB}:
\begin{eqnarray*}
\mathcal{E}_{SS}^{e\to p} & = & \frac{1}{\mu_{0}}\frac{e}{m_{e}}\frac{\mu_{0}}{4\pi}\langle\Psi|\int d^{3}x\ B_{p}(x)\cdot\left(-\Delta\frac{S_{e}(x_{e})}{|x-x_{e}|}\right)|\Psi\rangle\\
 & = & -\frac{e}{m_{p}}\frac{e}{m_{e}}\frac{\mu_{0}}{4\pi}\langle\Psi|\int d^{3}x\ \left(\nabla_{x}\times\left(\nabla_{x}\times\frac{S_{p}}{|x-x_{p}|}\right)\right)\cdot S_{e}\delta(x-x_{e})|\Psi\rangle\\
 & = & -\frac{e}{m_{p}}\frac{e}{m_{e}}\frac{\mu_{0}}{4\pi}\langle\Psi|\left[\left(\nabla_{x}\times\left(\nabla_{x}\times\frac{S_{p}}{|x-x_{p}|}\right)\right)\cdot S_{e}\right]_{x=x_{e}}|\Psi\rangle
\end{eqnarray*}
Through the vector calculus identities
\begin{eqnarray*}
-\left(\nabla_{x}\times\left(\nabla_{x}\times\varphi(x)S_{p}\right)\right)\cdot S_{e} & = & \left(S_{e}\times\nabla_{x}\right)\cdot\left(S_{p}\times\nabla_{x}\right)\varphi(x)\\
\left(S_{e}\times\nabla_{x}\right)\cdot\left(S_{p}\times\nabla_{x}\right)\frac{1}{|x-x_{p}|} & = & \left[\left(S_{e}\cdot S_{p}\right)\Delta_{x}-\left(S_{e}\cdot\nabla_{x}\right)\left(S_{p}\cdot\nabla_{x}\right)\right]\frac{1}{|x-x_{p}|}
\end{eqnarray*}
we obtain 
\begin{eqnarray*}
\mathcal{E}_{SS}^{e\to p} & = & \frac{e}{m_{p}}\frac{e}{m_{e}}\frac{\mu_{0}}{4\pi}\langle\Psi|\left[\left[\left(S_{e}\cdot S_{p}\right)\Delta_{x}-\left(S_{e}\cdot\nabla_{x}\right)\left(S_{p}\cdot\nabla_{x}\right)\right]\frac{1}{|x-x_{p}|}\right]_{x=x_{e}}|\Psi\rangle\\
\mathcal{E}_{SS} & = & \frac{e}{m_{p}}\frac{e}{m_{e}}\frac{\mu_{0}}{4\pi}\langle\Psi|\left[\left[\left(S_{e}\cdot S_{p}\right)\Delta_{x}-\left(S_{e}\cdot\nabla_{x}\right)\left(S_{p}\cdot\nabla_{x}\right)\right]\frac{1}{|x-x_{e}|}\right]_{x=x_{p}}|\Psi\rangle
\end{eqnarray*}
since both contributions are identical. The spin-spin interaction
is symmetric in front of the interchange of electron and proton and
then carries a factor one in front. 

\subsection{Interaction with an external electric field}

Suppose the atom is in interaction with an external electric field,
$E_{Z}.$ The electromagnetic interaction is then:
\[
{\cal E}_{AZ}=\epsilon_{0}\int dx\:\left((E_{A}+E_{A}^{am})\cdot E_{Z}\right)
\]
where $E_{A}=-\nabla V$ and $E_{A}^{am}=-\frac{\partial A}{\partial t}$
being $(V,A)$ the electromagnetic potentials associated to the atom:
\begin{eqnarray*}
A & = & \frac{\mu_{0}}{4\pi}\nabla\times\langle\Psi|\left(\frac{e}{m_{e}}\frac{M_{e}}{|x-x_{e}|}-\frac{e}{m_{p}}\frac{M_{p}}{|x-x_{p}|}\right)|\Psi\rangle\\
V & = & \frac{e}{4\pi\epsilon_{0}}\langle\Psi|\left(\frac{1}{|x-x_{e}|}-\frac{1}{|x-x_{p}|}\right)|\Psi\rangle
\end{eqnarray*}

\part*{Concluding remarks}

\subsection*{Quantum mechanics and instrumentalism}

The production of QM (a name attributed to Max Born \citep{born55-nobel})
under the leadership of the Copenhagen school was embraced in an instrumentalist
philosophical approach. Historically, the development was guided by
the correspondence with classical mechanics (although it deals almost
completely with electromagnetic interactions). Two of the three central
productive tools in the creation of QM are: imagination (as it is
illustrated by Bohr's well-known planetary atom) and analogy of forms,
i.e., the proposal of mathematical relations produced in a different
context and imported into QM without satisfying the conditions necessary
for their meaningful production. \footnote{For example: the Hamiltonian for a charged (point) particle used in
classical mechanics, as can be seen in \citep{pauli1927quantum,dirac1928quantum},
or Thomas' relativistic correction to the spin-orbit energy \citep{thoma26}
where, in order to patch Zeeman's formula obtained by analogy and
refuted by experiments, Thomas applies Lorentz transformations relating
two systems which are \emph{not} in uniform motion with constant velocity
with respect to each other (hence, the electromagnetic fields described
in each system are \emph{not} related by Lorentz transformations).} 

The third element is the adoption of instrumentalism. The Copenhagen
school was able to make important advances in QM introducing a matrix
formalism, but as Born confessed in his Nobel lecture: “What the real
significance of this formalism might be was, however, by no means
clear“ \citep{born55-nobel}. The situation is not a surprise since
instrumentalism is focused on (predictive) success rather than significance.
Its approach is more akin to technology than to science. It relies
on trial and error, success or failure in reproducing and collating
experimental results. The process of patching contradictions or inconsistencies
with ad-hoc hypotheses makes refutation almost impossible. In so doing,
it puts an undue pressure over the experimental side, since it requires
experiments each day more complex, scarce and in need of interpretation.
Compensating refutations by introducing ad-hoc patches leads more
often than not to specialism, since different patches need not be
mutually consistent. Seeking for the unity of knowledge moves precisely
in the opposite direction. The explosion of Science into multiple
sciences, each one constantly fragmenting in a fractal design, is
not a consequence of the complexity of nature but rather the consequence
of sacrificing reason in the altar of productivity.

The ``correspondence principle'', in the context of the utilitarian
development, is what remains of the idea of cognitive surpass (generalisation)
where the old becomes a particular instance of the new.

In terms of the correspondence with classical mechanics, the idea
of point particles is central to the statistical QM description. The
concept of particle is so strongly associated to the extreme idealisation
as points that, quite often, their point character is not mentioned
until late in the works. If particles in classical mechanics, when
possible, are idealised as mathematical points, in QM particles \emph{are}
mathematical points, the localisation in space becomes an essential
property of what in conventional (textbook) QM is considered a particle. 

Point particles are not an invention of the Copenhagen school, they
are present as idealisations in almost every textbook of physics.
In his theory of the electron, Lorentz \citep{lorentz1904electromagnetic}
considered the electron as a point surrounded by a cloud of electricity
located in the nearby ether. Further back in time, Faraday had opposed
these views in his theory of vibrating rays \citep[p. 447]{fara55}
indicating that neither the material point nor the ether were needed
if we considered matter as an inference arising as a consequence of
the observation of EM action. Faraday recommended extreme prudence
in introducing such inventions since in the past these metaphysical
elements had proven to be difficult to remove.

The scene changed with the irruption of the works by de Broglie \citep{debroglie23,de1924recherches}
and Schrödinger \citep{schrodinger1926undulatory} who introduced
the wave formulation of QM, although both of them relied in part on
point particles. Later, Schrödinger strongly opposed this view.

As mentioned in the Introduction, the EM interactions in Quantum Mechanics
were at variance with  electrodynamics. The expected radiation of
an electron orbiting the proton in an hydrogen-like atom was suppressed
by ruling that no such radiation occurs for some special orbits. In
short, conventional EM did not apply to quantum phenomena. The unity
of physics was broken as it was broken a few years earlier when Special
Relativity and its immaterial ether \citep{einstein24} was accepted
by the guild of physics as the cornerstone of electromagnetism, thus
breaking with Newtonian mechanics.

The self-appraisal of instrumentalism perceives these cognitive decisions
as realism or pragmatism and not as deficits. Instrumentalism advances
without retracing its steps, without reconsidering the possibility
of being in error, patching ``theories'' and narrowing their scope
as needed to avoid refutation. Actually, when the word ``physicist''
was coined \citep{yeo2003defining} the increasing specialisation
and lack of unity in the advances of physics was the main concern
for philosophers of Nature such as Faraday and Whewell.

The uncomfortable situation of the wave-particle duality forced by
Schrödinger and de Broglie is left behind by Max Born who introduces
the (now) standard interpretation of wave-functions as probability
waves. As Schrödinger points out, this appears as the only form of
reconciliation with point particles at hand, and it is the form in
which QM has been socially transmitted to several generations of physicists.
Adopting this interpretation forces us to write the wave-function
in a basis of functions, $|o\rangle$, where the operator $\hat{O}$
representing the measurable property is diagonal, $\hat{O}|o\rangle=o|o\rangle,\mbox{ with }o\in\mathbb{R}$,
so the ``matrix'' representing the density of states, $\langle o|\Psi\rangle\langle\Psi|o^{\prime}\rangle$,
has only positive values in the diagonal and zero otherwise. Since
two non commuting matrices (operators) cannot be simultaneously diagonalised,
it is not possible to obtain probabilities for their joint measurement.
The theory then ruled that this impossibility is an imposition of
Nature, although in deriving it we have only used our interpretation.
This difficulty in separating what comes from the subject (the observer)
and what from the object (the observed) has been central to the criticism
of natural science by Goethe, Faraday, Whewell and Peirce. It is the
attitude in front of the (possible) error what makes the most striking
difference between the instrumentalist approach and the philosophical
approach. The instrumentalist scientist resorts to patches, restrictions
and limitations. The final outcome is posed by S. Carroll as ``Physicists
do not understand Quantum Mechanics. Worse, they don't seem to want
to understand it'' \citep{carroll19}.

In a constructivistic perspective, theories are constructs and their
interpretations must be totally in accordance with the abduction process
employed in the construction stage, since abduction and interpretation
form the two components of the phenomenological map \citep{Solari22-phenomenologico}.
In the constructivistic sense, to ``understand a theory'' implies
to be able to construct it rationally starting from observations and
experimental information through a process rooted in previous knowledge
(which might require adjustments under the light of the new information)
and using synthetic judgements which are put to test by the consequences
of the theory. QM has not been reconstructed under the idea of being
a statistical theory. We have then the right to ask: where was the
statistics introduced when all our considerations have been about
the interactions of point particles? The resolution of this conflict
is simple for the instrumentalist, they tell us: ``the construction
is not part of science'' or even more strongly: science is not about
the reality but just a matter of economy of thought. 

The view of inferred matter as having identity by itself with forces
that abut around was the standard view used in the construction of
physics in the late XIX century \citep[{[529],}][]{maxw73} and propagates
there on, without being given reconsideration. For Einstein \citep{eins36}
the concept of material point is fundamental for Newton's mechanics
(not just an idealisation). In his work Einstein criticises Lorentz'
proposal of electrons extending in the space (with soft borders in
\citep[§33]{lorentz1892CorpsMouvants} and spherical in \citep[§8]{lorentz1904electromagnetic})
in the form:
\begin{quote}
The weakness of this theory lies in the fact that it tried to determine
the phenomena by a combination of partial differential equations (Maxwell's
field equations for empty space) and total differential equations
(equations of motion of points), which procedure was obviously unnatural.
The unsatisfactory part of the theory showed up externally by the
necessity of assuming finite dimensions for the particles in order
to prevent the electromagnetic field existing at their surfaces from
becoming infinitely great. The theory failed moreover to give any
explanation concerning the tremendous forces which hold the electric
charges on the individual particles. \citep{eins36}
\end{quote}
The criticism is faulty in several accounts. As we have shown in Part
I, a single mathematical entity, Hamilton's principle, is able to
produce the equations of Electromagnetism and mechanics. Einstein
qualifies Lorentz' electron theory as ``unsatisfactory'' for using
finite-size particles, since  for Einstein forces (electromagnetic
fields) exist in the surface of particles. Simply, he ignores Faraday,
and furthermore, there seems to be a requirement for particles being
mathematical points. Einstein gives no argument to justify the character
of ``unnatural'' for Lorentz' procedure, actually, ``unnatural''
appears to stand for ``alien to my scientific habit''. In contrast,
mathematical statements require proofs as we offer in Part I. The
argument concerning a need for a force to held together the electron
is equally faulty, it assumes that there are parts that need to be
held together, that the quantum entity is a composite in which one
part can act on another part. Actually, acquaintance with Maxwell's
arguments, make us to realise that the electrostatic energy computed
in electromagnetism requires the possibility of dividing the charge
in infinitesimal amounts. Something that can be considered a reasonable
limit in macroscopic physics (Maxwell's situation) but is ridiculous
when it comes to the electron. 

Maxwell's concept of science and our own concept differ radically
from Einstein's. For Maxwell and the present work, the construction
is a relevant part of the theory, for Einstein:
\begin{quote}
There is no inductive method which could lead to the fundamental concepts
of physics. Failure to understand this fact constituted the basic
philosophical error of so many investigators of the nineteenth century.
It was probably the reason why the molecular theory, and Maxwell's
theory were able to establish themselves only at a relatively late
date. Logical thinking is necessarily deductive; it is based upon
hypothetical concepts and axioms. How can we hope to choose the latter
in such a manner as to justify us in expecting success as a consequence?
\citep{eins36}.
\end{quote}
Einstein's utterly instrumentalist statements contrast with the legend
of his early reading of Kant's ``Critic of pure reason'' which elaborates
in the opposite direction through the concept of \emph{synthetic judgement}.
The final question posed was answered by Peirce's abduction (retroduction)\citep{peir94}
and the whole issue was considered in Whewell's philosophy of science\citep{whewell1860philosophy},
also here in contrasting terms.

In contrast, Einstein accepted without criticism that ``Everywhere
(including the interior of ponderable bodies) the seat of the field
is the empty space. \citep{eins36}'' (referring to the EM fields)
despite being conscious that the idea perpetuated the essence of the
ether \citep{einstein24}. 

The reason we emphasise Einstein's positions is that he has been socially
instituted as the ``most perfect scientist'', something ``known''
even by the illiterate. Actually, acceptance of his statement requires
the previous acceptance of the utilitarian view of science.

Concluding the historical revision, the adoption of an instrumentalist
approach weakens the demands put on the construction of theories and
by so doing it speeds up the process of delivering tools for the development
of technologies at the price of sacrificing understanding. 

\subsection*{The present approach}

In the present work we addressed the challenge of constructing QM
as a cognitive surpass of EM, in such a form that EM and QM can be
regarded as two instances of the same theory. The electromagnetic
formulation had to be previously unified not as an aggregation of
compatible formulae but rather through a unified deduction of the
formulae from a common principle. Such step was previously taken in
\citep{solari2022symmetries} where Maxwell equations and the continuity
equation all derive from a variational principle of minimal action,
actually, the same principle that is used to generate Lagrangian dynamics
and also the Lorentz force. In the presentation of relational electromagnetism
\citep{solari2022symmetries} Maxwell equations represent the propagation
of EM effects in regions outside its production sources. The sources
of EM action are described only as Lagrangian parameters that can
be identified as charge densities and current densities and satisfy
the continuity equation. However, the entity of such densities inside
what we intuitively call matter is not prescribed by electromagnetism.
Extending EM into the material region promises then to produce the
laws of QM.

In this endeavour we can continue to use the name ``particle'',
but understanding a quantum particle as an unitary entity characterised
by mass (which is the form of influencing through gravitational actions),
charge, current (charge in motion) and magnetisation (spin), the latter
ones being the attributes related to EM effects. Furthermore, we consider
charge and mass-densities to be proportional and currents to bear
the same proportion with momentum than the charge-to-mass relation.
These densities cannot be thought of in analogy of the densities resulting
from aggregates of matter since it is not possible to have a self-action
or an interaction with only a portion of the quantum particle \footnote{Readers familiar with Special Relativity (SR) may claim that there
is a requirement of simultaneity at different points that might violate
postulates of SR. However, SR displays self contradictions \citep{Solari22-phenomenologico}
prior to any claim. Its logical problems are usually made invisible
by the instrumentalist approach that persuades us not to examine the
correspondence of the construction and the interpretation. Experimental
results support our view as well, for example \citep{aspect1982experimental}. }.

The wave-function in our approach will then be associated to the distribution
of charge, currents and magnetisation. The need to use a wave-function
and not the densities directly is taken from de Broglie's argumentation
to make contact with experiments that relate frequency to energy involving
Planck's constant. We must highlight that when analysing a free particle,
the kinetic energy appears as the result of the above mentioned hypothesis
and the conservation of charge, resulting in the usual ${\displaystyle \frac{p^{2}}{2m}}$,
not as analogy or correspondence with classical mechanics but rather
to be consistent with electromagnetism.

The wave-function is an attribute that keeps track of the state of
the particle and is used primarily to determine the interaction (electromagnetic
or gravitational) with other particles. In turn, the state of the
particle is not just an attribute of itself but of its environment
as well. Such view may result odd to physicists and chemists while
it is the standard view in ecology. Thus, the problem of measurement
is incorporated from the beginning since measuring means to enter
in contact with a device.

Cognitive surpass means in this regard that the natural laws that
govern the interaction of electron and proton as spatially separated
entities must be the same as those governing their microscopic interaction,
the differences corresponding to the context. When the electron and
proton conform a new entity, the hydrogen atom, they act as a unity
and interact with the environment as an atom. This observation suggests
that the atomic wave-function may be factored as $|\Psi\rangle=|\psi_{r}\rangle|\psi_{CM}\rangle$
while when we consider electron and proton as not entangled, it factors
as $|\Psi\rangle=|\psi_{p}\rangle|\psi_{e}\rangle$, these two expressions
constituting two different limits of the total wave-function. This
simple argument allows us to construct, in Part II, the Hamiltonian
of the hydrogen atom without resource to analogies, empirical corrections
such as gyromagnetic factors or ``relativistic corrections'', i.e.,
patches inspired by SR needed to match experimental results.

We have shown that the classical laws of motion, Schrödinger's equation
and the laws of Electromagnetism all derive from the same principle
of minimal action. It is proper to say that they form a unit of knowledge.
There is no much surprise in this fact, the principle of minimal action
was introduced by Hamilton as form of collecting the system of ordinary
equations of Lagrangian mechanics in one equation in partial derivatives.
In turn Lagrange introduced the form today known as Lagrangian in
order to incorporate in one expression  those force laws given by
their effects (usually called constraints of motion) and those given
as interactions (the original realm of Newton's theory). Maxwell constructed
Electromagnetism guided by the least action principle\footnote{Actually, this is the mathematics about which Hertz said: ``Many
a man has thrown himself with zeal into the study of Maxwell's work,
and, even when he has not stumbled upon unwonted mathematical difficulties,
has nevertheless been compelled to abandon the hope of forming for
himself an altogether consistent conception of Maxwell's ideas. I
have fared no better myself.`` Hertz proceeds then to ``understand''
Maxwell's theory, actually to accept Maxwell's formulae resorting
to the ether. See \citep[p.20,][]{hert93}}. 

The particle is indissolubly associated to its localisation when it
is though as: ``the particle is that thing that lies there''. In
such regard, the vibration of Faraday's rays becomes dissociated from
the particle. It is not an action performed by nature but rather by
our way of knowing. In contrast, the ``problem'' of measurement
in QM is dissolved in our construction, actually, all what is needed
is a quantum mechanical description of the measurement device and
its (usually) electromagnetic form of interacting with the system
of interest. The universe cannot be suppressed in an idealisation,
its presence will be always in action, to the very least in the form
of time. If in addition, we allow ourselves to separate the inferred
(matter) from what is sensed (action) as it is done in the pair QM
and EM, we simply create a fantasy, an illusion, where poor Schrödinger's
cat is both dead and alive until we measure. Actually, under the present
view, the quantum mechanical entity will have associated time-dependent
electromagnetic fields in as much it has not reached a stationary
state. A state initially described as a superposition of stationary
states will evolve into an apparent stationary state if we agree to
ignore the relation with the electromagnetic energy whose space locus
has not longer an overlap with the space locus of the material entity.
Notice that in order to make this statement we have to include the
localisation between the attributes of the entity creating a second
entity, the photon, associated to the EM radiated energy.

In the present work we have advanced a unified action integral inspired
in de Broglie and Schrödinger intuitions as well as Maxwell's electrodynamics
(Lemma \ref{lem:Quantum} to Theorem \ref{thm:Classical}). Through
Hamilton's principle we obtained the equations of motion ruling microscopic
systems in accordance with classical mechanics and electromagnetism.
Further, we reformulated Hamilton's principle in terms of the (conserved)
energy, obtaining a slightly more general equation of Schrödinger
type (Lemma \ref{lem:Hamiltonian-form.}to Lemma \ref{lem:stationary}).
Finally we showed the connection between the possible form of the
variations and Hermitian operators (Lemma \ref{lem:variations}).
Hamilton's variational principle appears as central to the development,
since it separates circumstantial issues from ``laws of nature''.
The law is what remains invariant regardless of initial or final conditions
(or condition at any time) \footnote{The original concept of variation by Hamilton reads ``...any arbitrary
increment whatever, when we pass in thought from a system moving in
one way, to the same system moving in another, with the same dynamical
relations between the accelerations and positions of its points, but
with different initial data...''\citep{hamilton1834}}. The occurrence of Hamilton's principle is the result of successive
cognitive surpasses: from Newton equations to Lagrange equations and
next to Hamilton's principle in Mechanics. Hamilton's structure was
used by Maxwell to organise Electromagnetism \citep{maxw73} and by
Lorentz to reorganise Maxwell's electromagnetism in terms of the electromagnetic
Lagrangian \citep{lorentz1892CorpsMouvants} introducing his EM-force
from the application of this principle. It was later elevated to the
level of fundamental principle by Helmholtz because it systematically
produces the conservation of energy, and finally Poincaré \citep{poin13foundations}
considered it a fundamental principle as well. Hence, Hamilton's principle
was born as cognitive surpass and evolved as a guide for the construction
of physics, a history that situates it as the primary choice for an
attempt to unify theories.

In Part II, the quantum mechanics that emerges from electromagnetism
has been tested in the construction of the internal energies for the
Hydrogen atom producing outstanding results, superior to those of
standard quantum mechanics. Equal in precision but definitely superior
in consistence and consilience. Another insight gained is that the
moving spin reveals its role as the carrier of \emph{electric} polarisation
in the spin-orbit interaction.

The path followed in this construction is blocked for the standard
presentation of QM as extension from mechanics by the early assumption
of point particles. The usual form used to incorporate the Lorentz
force in the case of point particles was the constructive basis for
several authors such as Pauli \citep{pauli1927quantum} and Dirac
\citep{dirac1928quantum}. This led to the identification of the momentum
$p$ with the operator $-i\hbar\nabla$ (in all situations), a compensating
error sustained in Bohr's principle of correspondence. In turn, this
chain of decisions made impossible to associate the operator with
the current and prevented the production of EM interactions in QM
except by resource to analogy.

In contrast, in this work the unity of concepts is sought in the fundamentals.
The Lorentz force was integrated to Maxwell equations in \citep{solari2022symmetries}
(see Part I as well), without restrictions to point particles, using
Lorentz' Lagrangian formulation. The integration with mechanics was
then performed by incorporating the kinetic part of the action integral
in the form suggested by the QM of the isolated particle, thus reuniting
the \emph{material} and \emph{action} sides of the entity. After this
synthetic effort was completed, Hamilton's principle showed the extraordinary
power of consilience which was its initial justification \citep{hamilton1834}.
It is worth mentioning that so far there has been no need to request
the quantification of the EM field. All results are consistent with
Planck's view \citep{brush2002cautious} who recognised that the exchange
of energy in absorption and emission processes was quantified yet,
this fact does not imply that the field itself needs to be quantified
a priori, being the latter an idea usually attributed to Einstein
\citep{einstein1905photoelectric}. 

With the present work we have achieved the unification of the foundations
of classical mechanics, electromagnetism and quantum physics in a
sequence of works. This view is incompatible with the statistical
interpretation of QM where measuring is a sort of magic act in which
the laws of QM do not apply. As Carroll puts it ``Quantum Mechanics,
as it is enshrined in textbooks, seems to require separate rules for
how quantum objects behave when we’re not looking at them, and how
they behave when they are being observed'' \citep{carroll19}. Furthermore,
a condition for a theory to be considered rational consists in the
possibility of ``filling the gap'', which means that the details
left without specification can be provided in accordance to the theory.
The current QM is unable to specify the details of the measurement,
they remain in an abstract sphere where apparently, for example, the
position of the electron relative to the proton in the hydrogen atom
can be determined with as much precision as desired, yet not a word
is said on how such measurement would be performed. It fails then
a requisite of rationality as discussed in general form in \citep{sola23-razo-retro}.

We have shown that quantum mechanics can be conceived in a form that
does not oppose classical mechanics but coexists with it and complements
electromagnetism. The consilience of the present formulation not only
contrasts with the statistical interpretation of quantum mechanics,
it contrasts with what is implied in the notion of interpretation,
i.e., that equations are first herded together as needed in terms
of their usefulness without regard to their meaning, and a posteriori
meaning is attributed in the ``interpretation'' to promote acceptance
and facilitate the use of the equations. Such procedure is not needed,
it is not forced by the limitations of our mind and much less by Nature.
Any claim regarding the Truth in such collections is ludicrous. However,
students in sciences are indoctrinated constantly in terms of the
Truth value of such ``theories''. It is the social power of the
guild of physicists what prevents progress towards understanding.

There are two obstacles that have to be overcome to grasp our presentation,
the first is the entrenched belief in the representation of fundamental
quantum particles as points. Schrödinger was, in our limited knowledge,
the only one to point out this problem, yet he did not notice that
this view was threaded with Special Relativity (SR). The second obstacle
is the belief that Special Relativity (SR) is the cornerstone of electromagnetism.

Einstein's instrumentalism is declared in: 
\begin{quote}
Physics constitutes a logical system of thought which is in a state
of evolution, and whose basis cannot be obtained through distillation
by any inductive method from the experiences lived through, but which
can only be attained by free invention. The justification (truth content)
of the system rests in the proof of usefulness of the resulting theorems
on the basis of sense experiences, where the relations of the latter
to the former can only be comprehended intuitively. \citep[p. 381]{eins36}
\end{quote}
SR closed the debate around the ether at the price of elitism \citep{essen1978relativity,lerner2010big},
and dropping the advances made by Lorentz in terms of a Lagrangian
formulation of Maxwell's theory. Only Lorentz' law of forces was kept,
a law that was herded into electromagnetism probably to hide that
it comes from Galilean transformations, hence, its deduction by Lorentz
casts doubts on Einstein's alleged role of Lorentz transformations
\citep{einstein1907relativity} relating to inertial systems. We cannot
avoid wondering what ``inertial system'' means for Einstein, who
believes that Newton's mechanics rests upon absolute space \citep{einstein24}.
But the problem is not Einstein, all of us are limited and to show
him limited only proves what is not disputed: Einstein was human.
The force driving the social construction of idols is the rent obtained
by the clergy. In situations like this, the philosophical debate is
obscured and impeded by the social struggle.  Any one who has objected
the idea of SR as the cornerstone of EM has been threatened by the
guild of physics. We came to know only the testimony of those strong
enough to (socially) survive the attacks. Let us name them: Essen
\citep{essen1971special,essen1978relativity} who wrote ``The theory
is so rigidly held that young scientists dare not openly express their
doubts'' as abstract of his 1978 intervention. Essen provided evidence
that SR does not make sense in experimental terms. Dingle, warned
us about the epistemological problems of SR \citep{ding60}, the relevance
of the Doppler effect by itself \citep{ding60a} and gave testimony
of the attacks he endured \citep{ding72}. Phipps insisted from both
theoretical and experimental points of view in the relational view
of physics \citep{phipps2006old,phipps2014invariant}.

If we go back in time, the men in charge of producing theories about
nature were philosophers-mathematicians. For Galileo, mathematics
provided the language to understand the universe by the philosopher,
who was understood as a master at reasoning. When the production of
scientific results was ``industrialised'' a new social character
was born: the scientist (in particular, the physicist). The professional
scientist, produced at the universities, was declared free from the
control of philosophy (i.e., reason), each science had its method
and forms of validations, it was proclaimed. The guild of the profession
was then left in charge of the quality control. The immediate consequence
of this social change in physics was a new form of explanation that
left behind natural philosophy and empowered the ``second physicist''
\citep{jung17} with the social responsibility of producing theories.
The technical success of Electromagnetism contrasted with the failure
of the new epistemology that had made of the ether the weapon for
attacking the remains of the old epistemology, for example, e.g. Göttingen's
electromagnetism \citep{sola24-perdida}. When it became clear that
the ether was a no-end route, the task of saving the theoretical physicist
became urgent. The merit of SR is this social rescue and is also the
ultimate reason for not being challenged.

The present work, constructed much under the old epistemological view
shows that physics depends on epistemology, and through epistemology
it depends on social interests that have nothing to do with Nature.
The immediate utilitarian consequences appear as unchanged but the
foundations for advancing understanding of nature appear as dubious
if the instrumentalist perspective is adopted.

We believe we have shown by example that when we adopt a different
\emph{epistemic praxis} such as we have done in this work and in our
reconstruction of EM \citep{solari2022symmetries}, we reach a different
theoretical understanding, actually a deeper understanding when the
epistemic praxis is more demanding, as it is the relation between
our dualist phenomenology \citep{Solari22-phenomenologico} and instrumentalism.
Thus, physics depends not only on the observable natural phenomena
but it depends as well on our philosophical disposition. Yet, if the
concept of theory is weakened so as to be reduced to the equations
(cf. \citep[p. 21][]{hert93}), the distinctions between theories
mostly fade out, the phenomenological link disappears and interpretation
crops up to guide the use of the formulae after sacrificing the unity
of thought, critical and phenomenological actions.

In summary, there is no compelling reason to believe scientific results
that have been achieved by weakening the old scientific demand for
Truth into the current demand of usefulness. Newtonian physics is
not incompatible with EM, as it is usually preached; and both of them
can live in harmony with quantum mechanics. There is no reason for
abandoning (relational) Cartesian geometry or to consider time as
an ``odd'' spatial coordinate. There is no license to use formulae
outside their range of validity as it is too often done. 

Too often in the course of this investigation we have encountered
misrepresentations of the thoughts and writings of the natural philosophers
that developed physics until the middle of the XIX century, when the
\emph{scientist} was born. Abridged representations of their ideas
and writings creep in as soon as the creators fade out. Absolute space
is attributed to Newton's mechanics and ``true motion'' disappears
\citep{sola22}, Faraday becomes a supporter of the ether rather than
the careful philosopher he was, absolutely inclined to entertain doubts
as much as possible avoiding to precipitate into simplifying inventions.
Faraday writes in a letter to R Taylor:
\begin{quote}
But it is always safe and philosophic to distinguish, as much as is
in our power, fact from theory; the experience of past ages is sufficient
to show us the wisdom of such a course; and considering the constant
tendency of the mind to rest on an assumption, and, when it answers
every present purpose, to forget that it is an assumption, we ought
to remember that it, in such cases, becomes a prejudice, and inevitably
interferes, more or less, with a clear-sighted judgment {[}...{]}
\end{quote}
Maxwell's plead to consider the hypothesis of the ether as worth of
research \citep[p. 493][]{maxw73} was transformed as well into a
belief when the story was told. Yielding to the needs of the new social
position of science (and its emerging epistemic praxis), Maxwell's
theory was deprived of its mathematical construction lowering the
acceptance standard from correct into plausible. 

In our own experience, we have spent more time undoing promissory
hunches than constructing correct reasoning. It is worth noticing
that such hunches trigger exploratory actions and as such are important.
It is believing, instead of doubting them, what makes them prejudice.
What appears to us as correct is made of the debris of our errors.

The great minds that constructed Special Relativity and Quantum Theory
were passengers of their epoch, a time when ``conquering nature''
(the old imperialist dream of Francis Bacon) had finally taken prevalence
over the ``understanding nature,'' sought by his contemporary Galileo
Galilei. Our own work cannot escape the rule, even though we are not
``great minds''. We live in a time where Nature reminds us that
we must understand ki \textbf{}\footnote{The word \href{https://www.yesmagazine.org/issue/together-earth/2015/03/30/alternative-grammar-a-new-language-of-kinship}{ki}
has been proposed as a new and respectful pronoun for Nature and all
what is part of Nature. The word comes from Anishinaabe's language
but relates as well to the Japanese and Chinese Ki and to the old
English kin (like in kinship).} and, consequently, love ki. Science must then rescue the teaching
of the old masters, the natural philosophers, and their practice of
critical thinking.

\appendix

\section{\label{sec:Proof-of-variational}Proof of variational results }

\subsection{Proof of Lemma \ref{lem:action-final}:}
\begin{proof}
We use standard vector calculus operations on the integrand, namely
{\footnotesize{}
\begin{eqnarray*}
A\cdot j-\rho V & = & A\cdot\left(\frac{1}{\mu_{0}}\nabla\times B-\epsilon_{0}\frac{\partial E}{\partial t}\right)-\epsilon_{0}V\nabla\cdot E\\
 & = & \frac{1}{\mu_{0}}\left(|B|^{2}-\nabla\cdot\left(A\times B\right)\right)-\epsilon_{0}A\cdot\frac{\partial E}{\partial t}-\epsilon_{0}\nabla\cdot(VE)+\epsilon_{0}E\cdot\nabla V\\
 & = & \frac{1}{\mu_{0}}\left(|B|^{2}-\nabla\cdot\left(A\times B\right)\right)-\epsilon_{0}A\cdot\frac{\partial E}{\partial t}-\epsilon_{0}\nabla\cdot(VE)+\epsilon_{0}E\cdot\left(-E-\frac{\partial A}{\partial t}\right)\\
 & = & \frac{1}{\mu_{0}}|B|^{2}-\epsilon_{0}|E|^{2}-\nabla\cdot\left(\frac{1}{\mu_{0}}A\times B+\epsilon_{0}VE\right)-\epsilon_{0}\frac{\partial}{\partial t}\left(A\cdot E\right)
\end{eqnarray*}
}{\footnotesize\par}
\end{proof}

\subsection{Proof of Lemma \ref{lem:Quantum}:}
\begin{proof}
Since the wave-function admits the independent variations $\psi\mapsto\psi+\delta\psi$
and $\psi\mapsto\psi+i\delta\psi$ we have that 
\begin{eqnarray*}
\psi\frac{\delta{\cal A}}{\delta\psi}\delta\psi+\frac{\delta{\cal A}}{\delta\psi^{*}}\delta\psi^{*} & = & 0\\
\frac{\delta{\cal A}}{\delta\psi}\left(i\delta\psi\right)+\frac{\delta{\cal A}}{\delta\psi^{*}}\left(-i\delta\psi^{*}\right) & = & 0
\end{eqnarray*}
where $\delta\psi^{*}=(\delta\psi)^{*}$, and hence, it is sufficient
to consider the variations of $\psi$ and $\psi^{*}$ as independent
variations that must result in a null change of ${\cal A}$. By partial
integration we obtain $-i\hbar\dot{\psi}^{*}\psi=-i\hbar\left(\psi^{*}\psi\right)_{,t}+i\hbar\psi^{*}\dot{\psi}$
and the stationary action corresponds to 
\[
\delta{\cal A}_{QM}=\int_{t_{0}}^{t_{1}}ds\left\langle \delta\psi^{*}\left[i\hbar\dot{\psi}+\frac{\hbar^{2}}{2m}\Delta\psi\right]\right\rangle =0.
\]
\end{proof}

\subsection{Proof of Lemma \ref{lem:Prep}}
\begin{proof}
We act the variations piecewise, recalling that $\delta\psi=\epsilon\mu$
gives $\delta\dot{\psi}=\left(\epsilon\mu\right)_{,t}=\dot{\epsilon}\mu+\epsilon\dot{\mu}$,
\begin{eqnarray*}
\mathcal{A}_{1} & = & \frac{i\hbar}{2}\int_{t_{0}}^{t_{1}}ds\left\langle \psi^{*}\dot{\psi}-\dot{\psi}{}^{*}\psi\right\rangle \\
\delta\mathcal{A}_{1} & = & \frac{i\hbar}{2}\int_{t_{0}}^{t_{1}}\epsilon\ ds\left\langle \mu^{*}\dot{\psi}-\dot{\psi}{}^{*}\mu\right\rangle +\frac{i\hbar}{2}\int_{t_{0}}^{t_{1}}ds\left\langle \psi^{*}\left(\epsilon\mu\right)_{,t}-\left(\epsilon\mu\right)_{,t}{}^{*}\psi\right\rangle \\
 & = & \frac{i\hbar}{2}\int_{t_{0}}^{t_{1}}\epsilon\ ds\left\langle \mu^{*}\dot{\psi}-\dot{\psi}{}^{*}\mu\right\rangle -\frac{i\hbar}{2}\int_{t_{0}}^{t_{1}}\epsilon\ ds\left\langle \dot{\psi}^{*}\mu-\mu{}^{*}\dot{\psi}\right\rangle \\
 & = & \int_{t_{0}}^{t_{1}}\epsilon\ ds\left\langle \mu^{*}\left(i\hbar\dot{\psi}\right)+\left(i\hbar\dot{\psi}\right){}^{*}\mu\right\rangle 
\end{eqnarray*}
after a partial integration in time of the $\dot{\epsilon}$ contribution
(in the second line) that does not modify the variation. Partial integrations
in space using Gauss' theorem give
\begin{eqnarray*}
\mathcal{A}_{2} & = & \frac{\hbar^{2}}{2m}\int_{t_{0}}^{t_{1}}ds\left\langle \psi^{*}\Delta\psi\right\rangle \\
\delta\mathcal{A}_{2} & = & \frac{\hbar^{2}}{2m}\int_{t_{0}}^{t_{1}}\epsilon\ ds\left\langle \mu^{*}\Delta\psi+\psi^{*}\Delta\mu\right\rangle \\
 & = & \frac{\hbar^{2}}{2m}\int_{t_{0}}^{t_{1}}\epsilon\ ds\left\langle \mu^{*}\Delta\psi+\mu\Delta\psi^{*}\right\rangle .
\end{eqnarray*}
 Since the variation complies with Maxwell equations, after some partial
integrations, we have:{\scriptsize
\begin{eqnarray*}
\delta\frac{1}{2}\int_{t_{0}}^{t_{1}}ds\left\langle \frac{1}{\mu_{0}}B^{2}-\epsilon_{0}E^{2}\right\rangle  & = & \int_{t_{0}}^{t_{1}}ds\left\langle \frac{1}{\mu_{0}}\delta B_{m}\cdot B-\epsilon_{0}\delta E_{m}\cdot E\right\rangle \\
 & = & \int_{t_{0}}^{t_{1}}ds\left\langle \frac{1}{\mu_{0}}A\cdot\left(\nabla\times\delta B_{m}\right)-V\delta(\epsilon_{0}\nabla\cdot E_{m})-\epsilon_{0}A\cdot\delta E_{m,t}\right\rangle \\
 & = & \int_{t_{0}}^{t_{1}}ds\left\langle A\cdot\delta j_{m}-V\cdot\delta\rho_{m}\right\rangle 
\end{eqnarray*}
} where $\delta j_{m}=\epsilon\ts{J}_{1}+\dot{\epsilon}\ts{J}_{2}$
and $\delta\rho_{m}=\epsilon\ts{X}$. Consequently, 
\begin{eqnarray*}
\delta\frac{1}{2}\int_{t_{0}}^{t_{1}}ds\left\langle \frac{1}{\mu_{0}}B^{2}-\epsilon_{0}E^{2}\right\rangle  & = & \int_{t_{0}}^{t_{1}}ds\left\langle A\cdot\epsilon\ts{J}_{1}+\dot{\epsilon}A\cdot\ts{J}_{2}-V\epsilon\ts{X}\right\rangle \\
 & = & \int_{t_{0}}^{t_{1}}\epsilon\ ds\left\langle A\cdot\ts{J}_{1}-\left(A\cdot\ts{J}_{2}\right)_{,t}-V\ts{X}\right\rangle 
\end{eqnarray*}
after another partial integration in time.
\end{proof}

\subsection{Proof of Corollary \ref{cor:J2=00003DE}}
\begin{proof}
$\ts{J}_{2}=-\epsilon_{0}\ts{E}$ satisfies the assumption $\ts{X}+\nabla\cdot\ts{J}_{2}=0$
of Lemma \ref{lem:Prep}, as well as $\ts{J}_{2}+\nabla\times W$
does it, for any suitable $W$. However, setting $\ts{J}_{1}:=\ts{J}_{1}+\nabla\times W_{,t}$,
the vector $W$ does not appear in any of the equations of the Lemma.
Note that the precise form of $\ts{J}_{1}$ as a function of $\mu$
has not been prescribed so far. 
\end{proof}

\subsection{Proof of Theorem \ref{thm:Classical}}
\begin{proof}
Note that the column $\ts{X}$ in the table corresponds to propagating
$\delta\psi$ on the quantum charge density $\rho=\psi^{*}\psi$. 

For the ``time/energy'' entry, the variation of the action corresponds
to setting $\mu\equiv\dot{\psi}$ in Lemma \ref{lem:Prep}. Since
$\epsilon$ is any possible variation function, the only way to ensure
a stationary action is to let the space integral be identically zero,
namely: {\scriptsize
\[
\left\langle \mu^{*}\left(i\hbar\dot{\psi}\right)+\left(i\hbar\dot{\psi}\right)^{*}\mu\right\rangle +\frac{\hbar^{2}}{2m}\left\langle \mu^{*}\Delta\psi+\mu\Delta\psi^{*}\right\rangle +\left\langle A\cdot\ts{J}_{1}-\left(A\cdot\ts{J}_{2}\right)_{,t}-qV\ts{X}\right\rangle =0.
\]
} It follows that $\dot{\psi}^{*}\left(i\hbar\dot{\psi}\right)+\left(i\hbar\dot{\psi}\right){}^{*}\dot{\psi}=0$
and $\dot{\psi}^{*}\Delta\psi+\dot{\psi}\Delta\psi^{*}=(\psi^{*}\Delta\psi)_{,t}$,
by repeated application of Gauss' theorem. Letting $\ts{J}_{1}=-2\epsilon_{0}\ts{E}{}_{,t}-\ts{j}$,
we propose $\ts{j}=j_{,t}$ so that $\ts{X}_{,t}+\nabla\cdot\ts{j}=0$.
Since $\ts{X}=(\psi^{*}\psi)_{,t}$, we have that $\ts{J}_{2}=-\epsilon_{0}\ts{E}=j$
and $\ts{J}_{1}=j_{,t}$. The conditions of Lemma \ref{lem:Prep}
are satisfied and

\begin{eqnarray*}
A\cdot\delta j-V\delta\rho & = & \epsilon\left[A\cdot\ts{J}_{1}-\left(A\cdot\ts{J}_{2}\right)_{,t}-V\ts{X}\right]\\
 & = & \epsilon\left[A\cdot\left(\ts{J}_{1}-\ts{J}_{2,t}\right)-A_{,t}\cdot\epsilon_{0}\ts{E}-V\ts{X}\right]\\
 & = & -\epsilon\left[-\left(A_{,t}+\nabla V\right)\cdot\epsilon_{0}\ts{E}\right]\\
 & = & -\epsilon\left[E\cdot j\right]
\end{eqnarray*}
The latter expression is the time-derivative of the electromagnetic
energy:
\begin{eqnarray*}
\frac{1}{2}(\epsilon_{0}E^{2}+\frac{1}{\mu_{0}}B^{2})_{,t} & = & E\cdot\epsilon_{0}E_{,t}+\frac{1}{\mu_{0}}B\cdot B_{,t}\\
 & = & E\cdot\left(\frac{1}{\mu_{0}}\nabla\times B-j\right)-\frac{1}{\mu_{0}}B\cdot\left(\nabla\times E\right)\\
 & = & -E\cdot j+\frac{1}{\mu_{0}}\nabla\cdot\left(B\times E\right)
\end{eqnarray*}
By Gauss' theorem, the divergence vanishes when integrated and we
finally obtain the equation
\[
\left\langle \partial_{t}\left[-\frac{\hbar^{2}}{2m}\left(\psi^{*}\Delta\psi\right)+\frac{1}{2}\left(\frac{1}{\mu_{0}}B^{2}+\epsilon_{0}E^{2}\right)\right]\right\rangle =0
\]
expressing the conservation of energy. 

For the ``position/velocity'' entry in the table, in order to obtain
a stationary action integral, a similar argument about the independence
and arbitrariness of the real-valued time function $\epsilon$ forces
the following vector valued integral to be zero:{\small{}
\begin{eqnarray*}
\left\langle -i\psi^{*}\left(k\cdot x\right)\left(i\hbar\dot{\psi}\right)+\left(i\hbar\dot{\psi}\right)^{*}i\left(k\cdot x\right)\psi\right\rangle  & + & \frac{\hbar^{2}}{2m}\left\langle -i\psi^{*}\left(k\cdot x\right)\Delta\psi+i\left(k\cdot x\right)\psi\Delta\psi^{*}\right\rangle \\
 & + & \left\langle A\cdot\ts{J}_{1}+\left(A\cdot\ts{E}\epsilon_{0}\right)_{,t}-V\ts{X}\right\rangle 
\end{eqnarray*}
}where $\ts{X}=(ix\psi)^{*}\psi+\psi^{*}(ix\psi)=0$. Hence, we have
$-\ts{J}_{2}=\epsilon_{0}\ts{E}=0$. The continuity equation is satisfied
by taking $\ts{J}_{1}=0$. The electromagnetic integral is identically
zero and we obtain
\[
\left\langle k\cdot\left(-i\psi^{*}x\left(i\hbar\dot{\psi}\right)+\left(i\hbar\dot{\psi}\right){}^{*}ix\psi\right)\right\rangle +\frac{\hbar^{2}}{2m}\left\langle k\cdot\left(-i\psi^{*}x\Delta\psi+ix\psi\Delta\psi^{*}\right)\right\rangle =0.
\]
The first term gives $k\cdot\left(\hbar\partial_{t}\left\langle \psi^{*}x\psi\right\rangle \right)$
while the second reduces to $\frac{i\hbar^{2}}{2m}k\cdot\left\langle \psi^{*}\left(\nabla\psi\right)-\left(\nabla\psi\right){}^{*}\psi\right\rangle $
after repeated application of Gauss' theorem. Finally,
\[
0=\partial_{t}\left\langle \psi^{*}x\psi\right\rangle +\frac{i\hbar}{2m}\left\langle \psi^{*}\left(\nabla\psi\right)-\left(\nabla\psi\right){}^{*}\psi\right\rangle =\partial_{t}\left\langle \psi^{*}x\psi\right\rangle -\frac{1}{q}\langle j\rangle.
\]

For the ``displacement/momentum'' case, we compute $\ts{X}=r\cdot\nabla\rho$,
set $\ts{j}=\left(r\cdot\nabla\right)j$ and $\ts{J}_{2}=-\epsilon_{0}\ts{E}=-r\rho$.
Then $\ts{J}_{1}-\ts{J}_{2,t}=\epsilon_{0}\ts{E}_{,t}+\ts{j}$, which
computes to $\ts{J}_{1}-\ts{J}_{2,t}=r\rho_{,t}+(r\cdot\nabla)j=-r(\nabla\cdot j)+(r\cdot\nabla)j$
and further $-r(\nabla\cdot j)+(r\cdot\nabla)j=-\nabla\times\left(r\times j\right)$.
Hence, up to a global divergence vanishing by Gauss' theorem,
\begin{eqnarray*}
A\cdot\left(\ts{J}_{1}-\ts{J}_{2,t}\right)-A_{t}\cdot\ts{J}_{2}-V\ts{X} & = & A\cdot\left(-\nabla\times\left(r\times j\right)\right)+\left(A_{t}+\nabla V\right)\cdot\left(r\rho\right)\\
 & = & -A\cdot\nabla\times\left(r\times j\right)-E\cdot\left(r\rho\right)\\
 & = & -\left(r\times j\right)\cdot\left(\nabla\times A\right)-r\cdot\rho E\\
 & = & -r\cdot\left(\rho E+j\times B\right).
\end{eqnarray*}
Finally, the arbitrariness of $\epsilon$ demands {\footnotesize{}
\[
r\cdot\left\langle \left(\nabla\psi\right)^{*}\left(i\hbar\dot{\psi}\right)+\left(i\hbar\dot{\psi}\right)^{*}\nabla\psi\right\rangle +\frac{\hbar^{2}}{2m}r\cdot\left\langle \left(\nabla\psi^{*}\right)\Delta\psi+\left(\nabla\psi\right)\Delta\psi^{*}\right\rangle -r\cdot\left\langle \rho E+j\times B\right\rangle =0.
\]
}By repeated use of Gauss' theorem we obtain {\footnotesize{}$\left\langle \left(r\cdot\nabla\psi\right)^{*}\Delta\psi+\left(r\cdot\nabla\psi\right)\Delta\psi^{*}\right\rangle =0$}.
Since the directions $r$ are arbitrary, we have 
\[
i\hbar\left\langle \left(\nabla\psi\right)^{*}\dot{\psi}-\dot{\psi}^{*}\nabla\psi\right\rangle =\frac{m}{q}\partial_{t}\left\langle j\right\rangle =\left\langle \left(j\times B\right)+\rho E\right\rangle .
\]

For the ``rotation/angular momentum'' entry $\ts{X}=\Omega\cdot\left(x\times\nabla\rho\right)$,
$\ts{J}_{1}=\ts{j}=\left(\Omega\cdot\left(x\times\nabla\right)\right)j-\Omega\times j$
and we set $\ts{J}_{2}=-\left(\Omega\times x\rho\right)$ since $\nabla\cdot\ts{J}_{2}=-\Omega\cdot\left(x\times\nabla\rho\right)=-\nabla\cdot\left(\Omega\times x\rho\right)$.

The proposed variation $\delta\psi$ enters in the space integrals,
which specifically become 
\[
\Omega\cdot\left\langle i\hbar\left(\left(x\times\nabla\psi^{*}\right)\dot{\psi}-\left(x\times\nabla\psi\right)\dot{\psi}^{*}\right)+\frac{\hbar^{2}}{2m}\left(\left(x\times\nabla\psi^{*}\right)\Delta\psi+\left(x\times\nabla\psi\right)\Delta\psi^{*}\right)\right\rangle .
\]
The second term vanishes as in the previous case, while the first
term reads
\begin{eqnarray*}
\Omega\cdot\left\langle x\times\left(i\hbar\left\langle \left(\nabla\psi\right)^{*}\dot{\psi}-\dot{\psi}^{*}\nabla\psi\right\rangle \right)\right\rangle  & = & \frac{m}{q}\Omega\cdot\partial_{t}\left\langle x\times j\right\rangle 
\end{eqnarray*}
 The electromagnetic part reads
\begin{eqnarray*}
A\cdot\left(\ts{J}_{1}-\ts{J}_{2,t}\right)-A_{t}\cdot\ts{J}_{2}-V\ts{X} & = & A\cdot\left[\Omega\cdot\left(x\times\nabla\right)\right]j-\Omega\cdot\left(j\times A\right)\\
 & + & \Omega\cdot\left(\left(x\rho_{,t}\right)\times A\right)-\Omega\cdot\left[(x\rho)\times E\right]\\
 & = & -\Omega\cdot\left(x\times\left(j\times B\right)\right)-\Omega\cdot\left[x\times\rho E\right]\\
 & = & -\Omega\cdot\left(x\times\left(\rho E+j\times B\right)\right).
\end{eqnarray*}
Being $\Omega$ a vector of arbitrary direction, equating the variation
to zero we get
\[
\frac{m}{q}\partial_{t}\left\langle x\times j\right\rangle -\left\langle x\times\left(\rho E+j\times B\right)\right\rangle =0
\]
which is the evolution of the angular momentum under the action of
the Lorentz force.
\end{proof}

\subsection{Proof of Lemma \ref{lem:Hamiltonian-form.}}
\begin{proof}
We only need to check that $\left(\delta\rho\right)_{,t}+\nabla\cdot\delta j=0$
which follows from computing
\[
\ts{J}_{1}-\ts{J}_{2,t}=-\epsilon_{0}\ts{E}_{,t}-\ts{j}=-\frac{1}{\mu_{0}}\nabla\times\ts{B}
\]
according to Maxwell's equations. Substituting in the result of the
preparatory lemma and using the arbitrary $\epsilon(t)$ argument,
the results follow. 
\end{proof}

\subsection{Proof of Lemma \ref{lem:stationary}}
\begin{proof}
We begin from Eq. \ref{eq:Hamilton} of Lemma \ref{lem:Hamiltonian-form.}.
We write $\ts{X}=\left(\mu^{*}\psi+\mu\psi^{*}\right)$ and set $\ts{j}=-{\displaystyle \frac{i\hbar q}{2m}}\left(\mu^{*}\nabla\psi-\psi\nabla\mu^{*}+\psi^{*}\nabla\mu-\nabla\psi^{*}\mu\right)$.
According to Lemma \ref{lem:Prep} the variation of the electromagnetic
contribution to the action reads:
\[
\int_{t_{0}}^{t_{1}}\epsilon\ ds\left\langle A\cdot\ts{J}_{1}-\left(A\cdot\ts{J}_{2}\right)_{,t}-V\ts{X}\right\rangle 
\]
which according to Lemma \ref{lem:Hamiltonian-form.} is:
\[
\int_{t_{0}}^{t_{1}}\epsilon\ ds\left\langle -\frac{1}{\mu_{0}}A\cdot\left(\nabla\times\ts{B}\right)-\epsilon_{0}E\cdot\ts{E}\right\rangle 
\]

The variation of the action becomes: {\scriptsize
\[
\int\left(\epsilon(s)\,ds\right)\left\langle \mu^{*}\left(i\hbar\dot{\psi}\right)+\left(i\hbar\dot{\psi}\right)^{*}\mu\right\rangle +\frac{\hbar^{2}}{2m}\left\langle \mu^{*}\Delta\psi+\mu\Delta\psi^{*}\right\rangle -\left\langle -\frac{1}{\mu_{0}}A\cdot\left(\nabla\times\ts{B}\right)-\epsilon_{0}E\cdot\ts{E}\right\rangle .
\]
} Letting $\epsilon(t)$ to be constant, integrating in $-T\le s\le T$,
and letting $T\rightarrow\infty$, projects $\mu$ into its $\omega$
Fourier component, $\mu_{\omega}$, while the electromagnetic terms
integrate to $A\cdot\ts{j}_{\omega}-V\ts{X}_{\omega}$ projecting
in the same form the variation of density and current, with $\ts{X}_{\omega}=\left(\mu_{\omega}^{*}\psi_{\omega}+\mu_{\omega}\psi_{\omega}^{*}\right)$
and\\
 $\ts{j}_{\omega}=-{\displaystyle \frac{i\hbar q}{2m}}\left(\mu_{\omega}^{*}\nabla\psi_{\omega}-\psi_{\omega}\nabla\mu_{\omega}^{*}+\psi_{\omega}^{*}\nabla\mu_{\omega}-\nabla\psi_{\omega}^{*}\mu_{\omega}\right)$.
Since $\mu$ is arbitrary, its Fourier components are arbitrary and
the null variation of the action results in the requisite:
\[
\left\langle \mu_{\omega}^{*}\left(\hbar\omega\psi_{\omega}\right)\right\rangle +\frac{\hbar^{2}}{2m}\left\langle \mu_{\omega}^{*}\Delta\psi_{\omega}\right\rangle +\left\langle A\cdot\left(\frac{i\hbar q}{2m}\right)\left(\mu_{\omega}^{*}\nabla\psi_{\omega}-\psi_{\omega}\nabla\mu_{\omega}^{*}\right)-V\nu_{\omega}^{*}\psi_{\omega}\right\rangle =0.
\]
Partial integration gives $A\cdot\left(\nu^{*}\nabla\psi_{\omega}-\psi_{\omega}\nabla\nu^{*}\right)=A\cdot\left(2\nu^{*}\nabla\psi_{\omega}-\nabla\left(\psi_{\omega}\nu^{*}\right)\right)$
and by Gauss' theorem $\left\langle -A\cdot\nabla\left(\nu^{*}\psi_{w}\right)\right\rangle =\left\langle \nu^{*}\psi_{w}\nabla\cdot A\right\rangle $.
For an homogeneous magnetic field we have $A=\frac{1}{2}x\times B$
(and hence $\nabla\cdot A=0$) and further $A\cdot\left({\displaystyle \frac{i\hbar q}{2m}}\right)2\nu^{*}\nabla\psi_{\omega}=\nu^{*}\left({\displaystyle \frac{i\hbar q}{2m}}\right)\left(x\times B\right)\cdot\nabla\psi_{w}$,
which finally leads to

\[
\left\langle \nu^{*}\left(\hbar\omega\psi_{\omega}+\frac{\hbar^{2}}{2m}\Delta\psi_{\omega}-\frac{i\hbar q}{2m}B\cdot\left(x\times\nabla\psi_{\omega}\right)-qV\psi_{\omega}\right)\right\rangle =0.
\]
It follows that $\hbar\omega\psi\equiv\mathcal{E}\psi=-{\displaystyle \frac{\hbar^{2}}{2m}}\Delta\psi-{\displaystyle \frac{q}{2m}}B\cdot\left(x\times\left(-i\hbar\nabla\right)\psi\right)+qV\psi$.
\end{proof}

\subsection{Proof of Lemma \ref{lem:variations}}
\begin{proof}
Letting $\delta\psi=\epsilon\hat{O}\psi$, the operator $\hat{O}$
must be linear as a consequence of the infinitesimal character of
variations.  Acting with $\Theta$ on the continuity equation we get
\begin{eqnarray*}
\Theta\left(\rho_{,t}+\nabla\cdot j\right) & = & 0\\
\left(\Theta\rho\right)_{,t}+\Theta\left(\nabla\cdot j\right) & = & 0
\end{eqnarray*}
provided $\hat{O}_{,t}=0$. Since {\scriptsize
\[
\Theta\left(\nabla\cdot j\right)=\nabla\cdot\left(\Theta j\right)-{\displaystyle \frac{i\hbar q}{2m}}\left(\psi^{*}\left(\left[\nabla,\hat{O}\right]\cdot\nabla\right)\psi-\psi\left(\left[\nabla,\hat{O}\right]\cdot\nabla\psi\right)^{*}+\left(\left[\nabla,\hat{O}\right]\psi\right)^{*}\cdot\nabla\psi-\nabla\psi^{*}\cdot\left(\left[\nabla,\hat{O}\right]\psi\right)\right)
\]
} It is then possible to propose a suitable current $\ts{j}$ if
and only if {\scriptsize
\begin{equation}
-{\displaystyle \frac{i\hbar q}{2m}}\left(\psi^{*}\left(\left[\nabla,\hat{O}\right]\cdot\nabla\right)\psi-\psi\left(\left[\nabla,\hat{O}\right]\cdot\nabla\psi\right)^{*}+\left(\left[\nabla,\hat{O}\right]\psi\right)^{*}\cdot\nabla\psi-\nabla\psi^{*}\cdot\left(\left[\nabla,\hat{O}\right]\psi\right)\right)=\nabla\cdot K\label{eq:condition}
\end{equation}
} for some vector $K$. In such a case, $\ts{j}=\Theta j+K$, satisfies
$\left(\Theta\rho\right){}_{,t}+\nabla\cdot\ts{j}=0$.

Let us further investigate condition eq.\eqref{eq:condition}. Clearly,
it implies {\scriptsize
\[
-{\displaystyle \frac{i\hbar q}{2m}}\left\langle \left(\psi^{*}\left(\left[\nabla,\hat{O}\right]\cdot\nabla\right)\psi-\psi\left(\left[\nabla,\hat{O}\right]\cdot\nabla\psi\right)^{*}+\left(\left[\nabla,\hat{O}\right]\psi\right)^{*}\cdot\nabla\psi-\nabla\psi^{*}\cdot\left(\left[\nabla,\hat{O}\right]\psi\right)\right)\right\rangle =\left\langle \nabla\cdot K\right\rangle 
\]
} Introducing the Hermitian adjoint, the lhs can be rewritten as
follows:
\begin{eqnarray*}
\frac{i\hbar q}{2m}\left\langle \psi^{*}\left(\left[\Delta,\hat{O}\right]-\left[\Delta,\hat{O}\right]^{\dagger}\right)\psi\right\rangle  & = & \left\langle \nabla\cdot K\right\rangle .
\end{eqnarray*}
 Hence, if $\left[\Delta,\hat{O}\right]$ is Hermitian (or equivalently
if $\hat{O}^{\dagger}=-\hat{O}$, since $\Delta$ is itself Hermitian),
eq.\eqref{eq:condition} ensures that an adequate $K$ exists with
$\left\langle \nabla\cdot K\right\rangle =0$. Conversely, if a vector
field $K$ exists such that 
\[
\int d^{3}x\,\left(\nabla\cdot K\right)=\int_{\Sigma}K\cdot\widehat{d\Sigma}=0
\]
where the volume integral is taken as limit over all space and $\Sigma$
is the oriented surface bounding the volume of integration, then $\hat{O}$
is anti-Hermitian. 
\end{proof}

\section*{Acknowledgements}

Support from Kungliga Fysiografiska Sällskapet (LUND) and Svefum is
gratefully acknowledged.

\section*{Declaration of Interest}

The authors declare that there exists no actual or potential conflict
of interest including any financial, personal or other relationships
with other people or organizations within three years of beginning
the submitted work that could inappropriately influence, or be perceived
to influence, their work.

\bibliographystyle{authordate1}

\end{document}